\documentclass[12pt,english]{article}
\usepackage{amsmath}
\usepackage{array}
\usepackage{longtable}
\usepackage{amssymb}
\usepackage{mathrsfs}
\usepackage{graphicx}

\makeatletter

\catcode`@=12 \topmargin -0.5in \evensidemargin 0.0in
\usepackage{babel}
\makeatother
\newcommand {\be}{\begin{equation}}
\newcommand {\ee}{\end{equation}}
\newcommand {\ba}{\begin{eqnarray}}
\newcommand {\ea}{\end{eqnarray}}

\begin{document}
\thispagestyle{empty}
\begin{flushright}
IPM/P-2009/039\\
\end{flushright}

\mbox{} \vspace{0.75in}

\begin{center}\textbf{{\large Pseudo-Dirac Neutrino Scenario:}\\
Cosmic Neutrinos at Neutrino Telescopes}\\

 \vspace{0.5in} \textbf{$\textrm{Arman Esmaili}$\footnote{arman@mail.ipm.ir}}\\
 \vspace{0.2in} \textit{Department of Physics, Sharif University of Technology\\P.O.Box 11365-8639, Tehran, IRAN}\\
 \vspace{0.2in}\textit{School of Physics, Institute for Research in Fundamental Sciences (IPM)\\P.O.Box 19395-5531, Tehran, IRAN}\\
 \vspace{.55in}\end{center}

\baselineskip 18pt
\begin{abstract}

Within the ``pseudo-Dirac'' scenario for massive neutrinos the
existence of sterile neutrinos which are almost degenerate in mass
with the active ones is hypothesized. The presence of these
sterile neutrinos can affect the flavor composition of cosmic
neutrinos arriving at Earth after traveling large distances from
astrophysical objects. We examine the prospects of neutrino
telescopes such as IceCube to probe the very tiny mass squared
differences $10^{-12}$~eV$^2<\Delta m^2<10^{-19}$~eV$^2$, by
analyzing the ratio of $\mu$-track events to shower-like events.
Considering various sources of uncertainties which enter this
analysis, we examine the capability of neutrino telescopes to
verify the validity of the pseudo-Dirac neutrino scenario and
especially to discriminate it from the conventional scenario with
no sterile neutrino. We also discuss the robustness of our results
with respect to the uncertainties in the initial flavor ratio of
neutrinos at the source.

PACS numbers: 14.60.Pq; 13.15.+g; 95.85.Ry
\end{abstract}

\newpage

\section{Introduction}

Analyses of the data from reactor \cite{Apollonio:2002gd},
accelerator \cite{Aliu:2004sq}, atmospheric \cite{Ashie:2005ik}
and solar \cite{Ahmed:2003kj} neutrino experiments conclusively
demonstrate the oscillation of neutrino flavors. The results of
these experiments can be interpreted by two independent mass
squared differences (between three active neutrinos). From the
direct measurement of the invisible part of the decay width of $Z$
boson ({\it i.e.}, $\Gamma_{Z \to \nu_\alpha \bar{\nu}_\alpha}$),
the number of active neutrinos lighter than $M_Z/2$ found to be
$N_\nu=2.92\pm 0.06$ \cite{Achard:2003tx,pdg} and from the fit of
the LEP data to Standard Model prediction it found to be
$N_\nu=2.994\pm0.012$ \cite{Drees:2001xw,pdg}. Thus, if an extra
light neutrino exists, it should be a sterile neutrino (singlet
under the gauge symmetries of the Standard Model.) Historically
the strongest hint for the existence of sterile neutrino came from
the short baseline LSND experiment \cite{Aguilar:2001ty}. Data of
the LSND experiment, with neutrino energy $E_\nu\sim 30$~MeV and
baseline $\sim 30$~m, suggested a 3+1-scheme (active+sterile) with
the new mass squared difference $\Delta m^2 \sim
\mathcal{O}(1)$~eV$^2$. The LSND result has not been verified by
the MiniBooNE experiment \cite{AguilarArevalo:2007it} and
considerable efforts have gone into reconciling the null result of
MiniBooNE with the LSND data \cite{Farzan:2008zv}. All the other
data of the neutrino experiments can be interpreted by assuming
only three active massive neutrinos without any need to introduce
sterile neutrinos in the data analyses. However, sterile neutrinos
can still be present in the yet not probed regions of the
parameter space $(\Delta m^2,\theta)$. These regions correspond to
sterile neutrinos almost degenerate in mass with the active ones,
with very tiny mass differences $\Delta m^2 \ll \Delta m^2_{sol}$.
The scenario of degenerate sterile neutrinos, the so-called
``Pseudo-Dirac'' \footnote{The reason for this nomenclature will
be described in Sect.~\ref{formalism}} scenario, has been proposed
long time ago in \cite{pseudopast} and has been studied in the
literature extensively \cite{pseudorecent}.

The prospect for the existence of light sterile neutrinos with
masses nearly degenerate with the masses of active neutrinos is
motivated in many theoretical extensions of the Standard Model
\cite{motivation}. From the observational point of view, probing
very small $\Delta m^2$ between sterile and active neutrinos needs
very long baselines. Neutrinos coming from the Sun (which is the
farthest observed source of neutrinos with continuous emission)
set the bound $\Delta m^2\lesssim 10^{-12}$~eV$^2$ on the
active-sterile mass splitting. Bounds from other performed or
forthcoming experiments will be discussed in Sect.~\ref{bounds}.
In this paper we concentrate on the effects of almost degenerate
sterile  neutrinos on the expected flux of cosmic neutrinos coming
from astrophysical sources.

The new generation of km$^3$ scale neutrino telescopes give a
unique opportunity to probe the very tiny $\Delta m^2$ in
pseudo-Dirac scenario. The cosmic neutrinos from sources such as
GRBs \cite{GRB}, AGN \cite{AGN} and type Ib/c supernovae
\cite{SNR} travel large distances over $\sim 100$~Mpc before
arriving at neutrino telescopes in Earth. With such extremely long
baseline, tiny mass squared differences as small as $\Delta m^2
\sim 10^{-19} \;{\rm eV}^2 \left(E_\nu/100\; {\rm GeV} \right)$
can be probed. The idea of using neutrino telescopes to discover
the sterile neutrinos present in pseudo-Dirac scenario was
proposed in
\cite{Crocker:2001zs,Crocker:1999yw,Beacom:2003eu,Keranen:2003xd}.
In order to probe small values of $\Delta m^2$, it has been
suggested to look at distortions in the spectrum of $\nu_\mu$ from
supernovae remnants in the average distance of $\sim 1-8$~kpc in
\cite{Crocker:2001zs} and the spectrum of $\nu_\mu$ from Galactic
center in \cite{Crocker:1999yw}. The authors of
\cite{Beacom:2003eu,Keranen:2003xd} evaluate the effect of the
pseudo-Dirac neutrinos on the flavor composition of the cosmic
neutrinos; {\it i.e.}, the deviation of the
$F_{\nu_e}:F_{\nu_\mu}:F_{\nu_\tau}$ (where $F_{\nu_\alpha}$ is
the flux of $\nu_\alpha+\bar{\nu}_\alpha$ at Earth) from the
expected value $1:1:1$ in the absence of sterile neutrinos.

In the measurement of flavor composition of neutrinos in neutrino
telescopes, the uncertainties in the relevant parameters and
experimental limitations should be taken into account. For
example, there are uncertainties in the mixing parameters of
neutrinos and also in the spectrum of the arriving neutrinos.
Also, the current constructed or proposed neutrino telescopes,
AMANDA$/$IceCube \cite{icecube}, NEMO \cite{NEMO}, NESTOR
\cite{NESTOR}, ANTARES \cite{ANTARES} and KM3NET \cite{KM3NET}
cannot identify all three flavors of the active neutrinos. Ref.
\cite{Esmaili} considers these uncertainties and experimental
limitations in the analysis of the cosmic neutrinos in order to
extract mixing parameters and flavor composition of neutrinos at
the source. In this paper, by considering the aforementioned
uncertainties and experimental limitations, we investigate the
potential of neutrino telescopes in discovering the pseudo-Dirac
nature of neutrinos. The initial flavor ratio of the neutrinos at
the source can also be a source of uncertainty in the calculation
of event rates in neutrino telescopes. We discuss the robustness
of our result to this kind of uncertainty.

The paper is organized as follows. In sect.~\ref{formalism}, the
pseudo-Dirac scenario for massive neutrinos is reviewed. In
sect.~\ref{bounds}, the current bounds on $\Delta m^2$ from
various neutrino experiments are summarized; and in
sect.~\ref{cosmic} the effects of sterile neutrinos on the flavor
composition of cosmic neutrinos are discussed.
Sect.~\ref{detection} is devoted to the production mechanism of
neutrinos at the source and their detection processes in the
neutrino telescopes. Various sources of uncertainties that enter
the calculation of event rates in neutrino telescopes are
enumerated. Sect.\ref{result}, summarizes the results of the
present analysis on the capability of neutrino telescopes to
discriminate between pseudo-Dirac and conventional scenarios. A
summary of the results and the conclusions are given in
sect.~\ref{conclusion}.

\section{Pseudo-Dirac Scenario\label{formalism}}

A simple and economic way to generate mass for neutrinos in the SM
is to add right-handed (sterile) neutrinos to the matter content
of SM. In the presence of $N_s$ right-handed (sterile) fields
$\nu_{kR}$ ($k=1,\ldots,N_s$), we define the following column
matrix $\Psi$ of $N=3+N_s$ left-handed fields
\begin{equation}\Psi=\left(\nu_{eL}, \nu_{\mu L}, \nu_{\tau L},
(\nu_{1R})^C, \ldots, (\nu_{N_s R})^C\right)^T,\end{equation}
where ${\nu}^C=\mathcal{C}\bar{\nu}^T$ and $\mathcal{C}$ is the
charge conjugation operator. For Majorana neutrinos which we
consider here $(\nu_{\alpha R})^C=\bar{\nu}_{\alpha L}$. Here we
consider models with at most three sterile neutrinos ($N_s\leq3$).
In the basis $\Psi$, the generic mass term for neutrinos is
\begin{equation}
\mathcal{L}_m=-\frac{1}{2}\overline{{\Psi}^C}M\Psi+{\rm
H.c.},\end{equation}
The $(3+N_s)\times(3+N_s)$ mass matrix $M$ is of the following
form (after electroweak symmetry breaking)
\begin{equation}\label{massmatrix}
 M=\left(%
\begin{array}{cc}
  m_L & m_D^T \\
  m_D & m_R^\ast \\
\end{array}%
\right),\end{equation}where $m_D$ is the $N_s\times3$ Dirac mass
matrix and $m_L$ and $m_R$ are the $3\times3$ left-handed and
$N_s\times N_s$ right-handed Majorana mass matrices, respectively.
The non-vanishing elements of $m_L$ and $m_R$ violate lepton
numbers while by assigning the lepton number $+1$ to sterile
neutrinos, $m_D$ conserves this symmetry. The left-handed mass
matrix $m_L$ is not invariant under the SM gauge group $SU(2)_L$
and should be zero unless other new particles (such as a new Higgs
triplet) are present. The elements of $m_R$ can take a wide range
of values, it can be as large as the GUT scale $\sim 10^{15}$ GeV
which are preferred in {\it see-saw} mechanisms, or it can vanish
like $m_L$ as a result of new gauge symmetries such as $SU(2)_R$
\cite{Langacker:1998ut}. The case $m_L=m_R=0$ and $N_s=3$ results
in pure Dirac neutrinos. In this case the six Weyl neutrinos
decompose into three pairs of neutrinos with degenerate masses.
The active-sterile mixing angle in each pair is maximal
$\theta=\pi/4$, but the active neutrinos do not oscillate to their
sterile partners because $\Delta m^2_{s_j
a_j}=m^2_{s_j}-m^2_{a_j}=0$, where $m_{s_j}$ and $m_{a_j}$ are the
masses of sterile and active neutrinos in the $j$-th pair,
respectively. Here we are interested in the case $m_L,m_R\ll m_D$.
The non-zero but very small values of the elements of $m_L$ and
$m_R$ lift the degeneracy in mass at each pair. In this
``pseudo-Dirac'' scenario, active-sterile mixing angle in each
pair is $\theta \simeq \pi/4$ and active-sterile oscillation can
in principle occur due to very small but non-zero $\Delta
m^2_{sa}$. To illustrate this point, let us consider the one
generation example. In this case, the mass matrices $m_L$, $m_R$
and $m_D$ in Eq.~(\ref{massmatrix}) are numbers (we assume that
all the masses are real.) In the pseudo-Dirac limit, we obtain
$\tan (2\theta)=|2m_D/(m_R-m_L)|\gg1$ and $\Delta m^2_{sa}\simeq
2m_D(m_L+m_R)\ll m_D$. Notice that in the pseudo-Dirac scenario,
neutrinos oscillate even in one generation, in contrast to pure
Dirac scenario where oscillation occurs only between generations.

In general the $N\times N$ symmetric mass matrix $M$ (where
$N=3+N_s$) can be diagonalized by $V_\nu^T MV_\nu=M_{diag}$, where
$V_\nu$ is a $N\times N$ unitary matrix. We choose the elements of
$V_\nu$ such that $M_{diag}={\rm
diag}(m_{a_1},m_{a_2},m_{a_3},m_{s_1},\ldots,m_{s_{N_s}})$. The
mixing matrix $V$ appearing in the weak charged-current
$J_W^\mu=2\overline{\Psi}_iV^\dagger \gamma^\mu l_{\alpha L}$ is a
$3\times N$ rectangular matrix with the elements $V_{\alpha
k}=\sum_{\beta=e,\mu,\tau}(V_l^\dagger)_{\alpha\beta}
(V_\nu)_{\beta k}$, where $V_l$ is the $3\times 3$ diagonalizing
unitary matrix of charged leptons mass matrix. In the case
$N_s=3$, the $6\times 6$ matrix $V$ can be parameterized by 12
mixing angles and 12 CP-violating phases (7 Dirac phases+5
Majorana phases.) It has been shown in \cite{Kobayashi:2000md}
that in the pseudo-Dirac limit $m_L,m_R\ll m_D$ and at first order
of perturbation in the small parameters $m_L/m_D$ and $m_R/m_D$,
the mixing matrix $V$ has only three mixing angles (responsible
for oscillation between the pairs) and three CP-violating phases
(1 Dirac phases+2 Majorana phases). This fact can be seen from the
explicit form of the matrix $V_\nu$ which diagonalizes the mass
matrix $M$ (in the pseudo-Dirac limit) \cite{Kobayashi:2000md}:

\begin{equation} \label{vnu}
V_\nu=\left(%
\begin{array}{cc}
  U_{PMNS} & 0 \\
  0 & U_R \\
\end{array}%
\right).\frac{1}{\sqrt{2}}\left(%
\begin{array}{cc}
  I_{3\times 3} & iX_{3\times N_s} \\
  (X_{3\times N_s})^T & -iI_{N_s\times N_s} \\
\end{array}%
\right),\end{equation} \\ where $U_{PMNS}$ is the $3\times 3$
conventional neutrino mixing matrix of left-handed neutrinos,
$U_R$ is the $N_s \times N_s$ unitary matrix which diagonalizes
the right-handed Majorana mass matrix, $I_{n\times n}$ is the
$n\times n$ identity matrix and the matrices $X_{3\times N_s}$
($N_s\leq3$) are:

\begin{equation}
X_{3\times 1}=\left(%
\begin{array}{c}
  1 \\
  0 \\
  0 \\
\end{array}%
\right),\qquad X_{3\times 2}=\left(%
\begin{array}{cc}
  1 & 0 \\
  0 & 1 \\
  0 & 0 \\
\end{array}%
\right),\qquad X_{3\times 3}=I_{3\times 3}. \end{equation}\\ The
flavor conversion probability between the active neutrinos
$P_{\alpha\beta}\equiv P_{\nu_\alpha \to \nu_\beta}(L,E_\nu)$ is

\begin{equation} P_{\alpha\beta}=\left|\left(V_\nu
\exp\left\{i\frac{M^2_{diag}L}{2E_\nu}\right\}V^\dagger_\nu\right)_{\alpha\beta}\right|^2.
\end{equation}\\ Using the explicit form of the matrix $V_\nu$ in
Eq.~(\ref{vnu}), the probability $P_{\alpha\beta}$ becomes

\begin{equation}\label{probability} P_{\alpha\beta}=\frac{1}{4}
\left| \sum_{j=1}^{3}U_{\alpha
j}\left\{e^{i(m^+_j)^2L/2E_\nu}+e^{i(m^-_j)^2L/2E_\nu}\right\}U_{\beta
j}^\ast\right|^2,\end{equation} where $m_j^+$ and $m_j^-$ are the
mass eigenvalues in the $j$-th pair of active and sterile
neutrinos; $U_{\alpha j}$ and $U_{\beta j}$ are the elements of
the $3\times 3$ mxixng matrix $U_{PMNS}$. Notice that this
relation reduces to the standard flavor conversion probability
formula in the limit of pure Dirac neutrinos $m_j^+=m_j^-$
($j=1,2,3$). By setting $m_j^+=m_j^-$ for the active neutrino
generations which do not have sterile partners,
Eq.~(\ref{probability}) also applies to cases with $N_s<3$.

Using Eq.~(\ref{probability}) in analyzing the data of oscillation
experiments gives information on $(m^+_j)^2-(m^-_j)^2$ in each
pair. In subsect.~\ref{bounds} we review the current bounds on
active-sterile mass square differences and the prospect of future
experiments to improve these bounds. In subsect.~\ref{cosmic} we
discuss the implications of Eq.~(\ref{probability}) on the flavor
composition of cosmic neutrinos.

\subsection{Current Bounds on $\Delta m^2$ and Sensitivity of
Future Experiments \label{bounds}}

An oscillation experiment with baseline $L$ and neutrino energy
$E_\nu$ can probe mass square difference $\Delta m^2\sim
E_\nu/(4\pi L)$. If $\Delta m^2\ll E_\nu/(4\pi L)$, the baseline
is too short for flavor oscillation to take place; on the other
hand, if $\Delta m^2\gg E_\nu/(4\pi L)$ so many oscillations take
place during the propagation and the oscillatory term should be
averaged out. In both of these cases it is not possible to derive
the value of $\Delta m^2$ in oscillation experiments.

Solar neutrino experiments with the baseline
1~AU~$\approx1.5\times10^{11}$~m and neutrino energy $E_\nu \sim
0.1-10$~MeV, can probe mass squared differences $\Delta
m^2\sim10^{-10}-10^{-12}$~eV$^2$. These very small values of
$\Delta m^2$ has been favored by the so-called ``Vacuum
Oscillation Solution'' of the solar neutrino problem, but as is
well-known this solution has been ruled out by KamLand
\cite{Apollonio:2002gd}. However, the sterile-active oscillation
with mass square differences $\Delta m^2\lesssim10^{-12}$~eV$^2$
can still be present as a subdominant effect in solar data. The
recent work \cite{deGouvea:2009fp} updates the solar data and
obtains $\Delta m^2 < 1.8 \times 10^{-12}$~eV$^2$ (at $3\sigma$
level) for the sterile-active mass splitting. This bound is the
most stringent bound on $\Delta m^2$. The flavor composition of
the neutrinos from core-collapse supernovae (SNe) also can change
due to an active-sterile oscillation from the SN to Earth. The
mean energy of the neutrinos from a SN explosion is $E_\nu \sim
30$~MeV. Thus, a SN explosion at a distance of $\sim 10$~kpc can
probe $\Delta m^2 \sim 10^{-19}$~eV$^2$. The constraint from the
data of the SN1987A data is not restrictive because of the low
statistics and high uncertainties in the mechanism of SNe
explosion \cite{Nunokawa:1997ct}. Construction of future Mton
water-\v{C}erenkov detectors can dramatically improve the current
bound or find a hint for sterile neutrinos hypothesizes in
pseudo-Dirac scenario \cite{Cirelli:2004cz}.

Population of the sterile neutrinos in the early universe and
their effects on the Big Bang Nucleosynthesis (BBN) can change the
abundance of light elements. The standard BBN, given the number of
the relativistic particles $N_\nu$ and the baryon asymmetry
$\eta=n_B/n_\gamma$, predicts the abundance of the light nuclei in
the universe. Assuming that the sterile neutrinos are produced
only in the active/sterile oscillation and the initial abundance
of sterile neutrinos at temperatures $T\gg$~MeV is zero, the
tightest limit comes from the ${}^4$He abundance: $\Delta m^2
\lesssim 10^{-8}$~eV$^2$ \cite{Cirelli:2004cz,Enqvist:1991qj}.

Two main non-oscillation neutrino experiments which probe neutrino
masses kinematically are tritium beta decay and neutrinoless
double beta decay ($0\nu\beta\beta$) experiments. Among them, the
$0\nu\beta\beta$ decay is sensitive to the Majorana or Dirac
nature of neutrinos. The rate of $0\nu\beta\beta$ decay is
proportional to the effective mass of the electron neutrino which
is defined as

\begin{equation}\label{mee} \langle m_{ee}\rangle=\left| \sum_{j=1}^{6} (V_\nu)^2_{ej}
m_j
\right| =\left| \sum_{j=1}^3
\left(\frac{U_{ej}}{\sqrt{2}}\right)^2(m^+_j-m_j^-)\right|\end{equation}\\
As mentioned after Eq.~(\ref{massmatrix}), in the limit
$m_L=m_R=0$ (pure Dirac neutrino) each Dirac neutrino is the
superposition of two Majorana neutrinos with degenerate masses and
opposite CP eigenvalues. It is easy to see that the Majorana
neutrinos in each pair interfere destructively in Eq.~(\ref{mee})
($m_j^+=m_j^-$) which results in $\langle m_{ee}\rangle=0$ for
pure Dirac neutrinos. In the pseudo-Dirac scenario with non-zero
Majorana masses and $m_L,m_R\ll m_D$, the cancelation is not exact
and $\langle m_{ee}\rangle\neq0$ but it is very small
\cite{Doi:1980yb}. Thus, it seems that the observation of a
positive signal in the next generation $0\nu\beta\beta$
experiments, with sensitivities $\langle m_{ee}\rangle\sim10$~meV,
will rule out the small values of $m_L$ and $m_R$ in the mass
matrix of neutrinos and therefore pseudo-Dirac scenario. However,
two points should be considered. The first one is that the value
of $\langle m_{ee}\rangle$ can still be significant if only one or
two families of neutrinos have sterile partners. Contribution of
each family to the value of $\langle m_{ee}\rangle$ depends on the
corresponding mixing matrix element $U_{\alpha j}$. This means
that, because of the small value of $U_{e3}$ ($\leq0.041$),
presence or absence of a sterile neutrino with a mass almost
degenerate with $\nu_{3L}$ do not change the value of $\langle
m_{ee}\rangle$ substantially; but the case with two sterile
neutrinos with a masses degenerate with $\nu_{1L}$ and $\nu_{2L}$
leads to an effective mass $\langle m_{ee}\rangle$ much smaller
than its value in the absence of sterile neutrinos. The second
point is that the dominant contribution to $0\nu\beta\beta$ decay
can come from new particles or physics beyond the SM, such as a
$V+A$ interaction \cite{Vergados:2002pv}. It is shown in
\cite{Schechter:1981bd} that a non-zero $0\nu\beta\beta$ decay
rate generates small $m_L$ through radiative corrections, which
results in pseudo-Dirac scenario for neutrino masses. Considering
these points, it is not easy to draw a conclusion on pseudo-Dirac
scenario from the results of the $0\nu\beta\beta$ experiments.

\subsection{Cosmic Neutrinos \label{cosmic}}

Neutrinos arriving at neutrino telescopes from astrophysical
sources travel distances of the order $L\sim 100$ Mpc. The flavor
conversion probabilities over these large distances can be
obtained by averaging out the oscillatory terms in
Eq.~(\ref{probability}). Two different scales of $\Delta m^2$ are
involved in Eq.~(\ref{probability}), one is the atmospheric,
$\Delta m^2_{atm}\sim10^{-3}$ eV$^2$, and solar, $\Delta
m^2_{sol}\sim10^{-5}$ eV$^2$, and the other is the very small
$\Delta m^2$ between the mass eigenstates in each pair of active
and sterile neutrinos. As it is shown in \cite{Farzan:2008eg} the
oscillatory terms depending on $\Delta m^2_{atm}$ and $\Delta
m^2_{sol}$ should be completely averaged out over these large
distances. The mass squared difference $\Delta m^2$ that can be
probed by neutrinos with energy $E_\nu$ which propagate through
distance $L$ is

\begin{equation} \frac{\Delta m^2}{{\rm eV}^2}=10^{-16}
\left(\frac{{\rm Mpc}}{L}\right)\left(\frac{E_\nu}{{\rm
TeV}}\right).\end{equation}\\ Thus, even for $L$ as large as
10~Mpc oscillatory terms given by $\Delta m^2\sim 10^{-17}$ eV$^2$
do not average out. After averaging out $\Delta m^2_{atm}$ and
$\Delta m^2_{sol}$, Eq.~(\ref{probability}) becomes
\cite{Beacom:2003eu}

\begin{equation} P_{\alpha\beta}=\sum_{j=1}^3 |U_{\alpha j}|^2
|U_{\beta j}|^2 \cos^2\left(\frac{\Delta m_j^2
L}{4E_\nu}\right),\end{equation}\\ where $\Delta
m_j^2\equiv(m^+_j)^2-(m^-_j)^2$ is the mass squared difference in
the $j$-th pair. Thus, if the initial flavor composition of
neutrinos in the source is $w_e:w_\mu:w_\tau$, the flavor
composition of the neutrino beam arriving at Earth will be
$F_{\nu_e}:F_{\nu_\mu}:F_{\nu_\tau}$, where

\begin{equation}
\label{flavor} F_{\nu_\alpha}=\sum_\beta w_\beta \sum_{j=1}^3
|U_{\alpha j}|^2|U_{\beta j}|^2 \cos^2 \left(\frac{\Delta m^2_j
L}{4E_\nu}\right).\end{equation}\\ The average of the cosine
factor for $\Delta m_j^2 L/4E_\nu\gg1$ is $1/2$. Thus, if for all
three pairs ($j=1,2,3$) the condition $\Delta m_j^2 L/4E_\nu\gg1$
is satisfied, all three $F_{\nu_\alpha}$ in Eq.~(\ref{flavor}) are
multiplied by $1/2$ such that the flavor ratios
$F_{\nu_e}:F_{\nu_\mu}:F_{\nu_\tau}$ do not change with respect to
the flavor ratios in the pure Dirac case $\Delta m_j^2=0$. In this
situation the only difference between the pure Dirac and
pseudo-Dirac scenarios is that the number of neutrinos arriving at
Earth is reduced by half in the pseudo-Dirac case. Thus, it is
very hard to verify pseudo-Dirac scenario for distances or mass
squared differences where $\Delta m_j^2 L/4E_\nu\gg1$ for
$j=1,2,3$. That is because the estimation of the overall
normalization of neutrino flux needs knowledge about the details
of the neutrino production mechanism in the source which is not
well understood.

However, the pseudo-Dirac scenario of neutrinos can be tested in
neutrino telescopes for the cases that only one or two sterile
neutrinos exist ($N_s<3$) or the distance of the source or the
mass square differences are such that all the three oscillatory
terms given by $\Delta m_j^2 L/4E_\nu$ do not average out. In
these cases the average of one or two of the cosine factors in
Eq.~(\ref{flavor}) are $1/2$ and the other cosine factors can be
replaced by $1$ (we do not consider the special situations where
$\Delta m_j^2 L/4E_\nu\sim1$). Thus, measuring the deviations of
the flavor ratios in Eq.~(\ref{flavor}) from their standard values
(in the absence of nearly degenerate sterile neutrinos) in
neutrino telescopes can shed light on these cases. In the next
section we discuss the details of the detection processes in the
neutrino telescopes and the feasibility of identifying different
neutrino flavors in these experiments. By taking into account the
realistic measurable quantities in neutrino telescopes and the
uncertainties in these measurements, we discuss to what extent it
is possible to measure the flavor ratio of cosmic neutrinos and
their deviations from the standard values.

\section{Detection and Production Processes\label{detection}}

In this section we briefly discuss neutrino flavor identification
in the km$^3$ scale neutrino telescopes such as IceCube or its
counterparts in the Mediterranean sea; KM3NET, NEMO, NESTOR and
ANTARES. A detailed description of the flavor tagging efficiencies
in a typical neutrino telescope can be found in
\cite{Esmaili,Beacom:2003nh}. Here we summarize the main points
relevant for the present analysis. Particularly, for the first
time, we take into account different sources of uncertainties in
the calculation of event rates and also a more realistic analysis
of the detectable events, such as the contribution of $\nu_\tau$
($\bar{\nu}_\tau$) to the $\mu$-tracks.

The flux of neutrinos arriving at Earth can come from a single
luminous point source or as a diffuse flux from sum over different
sources at different distances. The advantage of the point sources
is that the short period of the burst and the direction of
incoming neutrinos can be used to reduce the background events
(especially when the source can be identified using a different
method such as gamma photons for GRBs). Point sources with an
intense neutrino flux detectable at km$^3$-scale neutrino
telescopes can take place in the close-by galaxies located at a
distance of $\lesssim 10$~Mpc and such a source of neutrinos
yields about a few hundred neutrino events in IceCube
\cite{optimistic}.

Discriminating between different flavors of neutrinos is a great
challenge for neutrino telescopes. Two types of events are
completely distinguishable in the next generation of these
experiments: $\mu$-track events and shower-like events. Charged
Current (CC) and Neutral Current (NC) interactions of different
flavors can contribute to each of these events. $\mu$-track
events, which are the \v{C}erenkov light radiated by muons
propagating through the volume of the detector, get contributions
from two sources: i) $\mu$ ($\bar{\mu}$) produced in the CC
interaction of $\nu_\mu$ ($\bar{\nu}_\mu$); ii) CC interaction of
$\nu_\tau$ ($\bar{\nu}_\tau$) which produce $\tau$ ($\bar{\tau}$)
leptons and the subsequent {\it leptonic} decay of tau leptons
$\tau\to \mu\bar{\nu}_\mu\nu_\tau$ ($\bar{\tau}\to
\bar{\mu}\nu_\mu\bar{\nu}_\tau$) produce $\mu$ ($\bar{\mu}$).
Shower-like events have three sources: i) NC interactions of all
the three flavors of neutrinos; ii) CC interactions of $\nu_e$ and
$\bar{\nu}_e$; and iii) CC interaction of $\nu_\tau$
($\bar{\nu}_\tau$) which produce $\tau$ ($\bar{\tau}$) leptons and
their subsequent {\it hadronic} decays. The exact formulae for
calculating the rate for each of these events can be found in
Sect.~2 of \cite{Esmaili}.

The threshold energy of the detection of $\mu$-tracks and showers
in experiments such as IceCube, respectively, is $E^{th}_\mu\sim
100$ GeV and $E^{th}_{shower}\sim 1$ TeV \cite{icecube}. On the
other hand, the mean free path of the neutrinos with energy
$E^{cut}_\nu\sim 100$ TeV becomes of the order of $\sim
2R_{\oplus}$, the diameter of Earth. The exact values of the
$E^{th}_\mu$, $E^{th}_{shower}$ and $E^{cut}_\nu$ depend on the
details of the experiments, such as the geometry of the
photomultipliers in the volume of the detector and the direction
of the incoming neutrinos; and a dedicated analysis can be done
for each experiment. Here, in order to avoid considering the
absorption of neutrinos in the Earth, we restrict the analysis to
100 GeV $<E_\nu<100$ TeV.

The realistic quantity that can be measured in neutrino telescopes
is

\begin{equation} \label{ratioR} R=\frac{\text {Number of
Muon-track events}}{\text {Number of Shower-like
events}}.\end{equation}\\ The value of $R$ can be calculated if we
know the initial flux and the flavor ratio of neutrinos at the
source. The main mechanism of neutrino production at the
astrophysical sources is the interaction of the accelerated proton
by the ambient protons and photons. The decay of the secondary
particles ($\pi^\pm$, $K^\pm$, $D$, \ldots) produced in these $pp$
and $p\gamma$ interactions generate neutrinos and muons. For
example, the pion chain decays
$\pi^+\to\mu^+\nu_\mu\to(e^+\nu_e\bar{\nu}_\mu)\nu_\mu$ and
$\pi^-\to\mu^-\bar{\nu}_\mu\to(e^-\bar{\nu}_e\nu_\mu)\bar{\nu}_\mu$
generate neutrinos with the flavor ratio
$(\nu_\mu+\bar{\nu}_\mu)/(\nu_e+\bar{\nu}_e)\simeq2$. The exact
value of the flavor ratio $w_e:w_\mu:w_\tau$ depends on the
spectrum of the parent particles and the properties of the
production medium. A class of models based on the Fermi
acceleration mechanism for the particle in the source, predict a
power-law spectrum for neutrinos

\begin{equation} \label{powerlaw}
\frac{dF_{\nu_\beta}}{dE_{\nu_\beta}}=
\mathcal{N_{\nu_\beta}}E^{-\alpha}_{\nu_\beta}, \end{equation}\\
where $\alpha$ is the spectral index and $\mathcal{N_{\nu_\beta}}$
is a normalization factor. Acceleration of particles through Fermi
acceleration mechanism \cite{fermiacc} results in $\alpha=2$ for
neutrino spectrum. However, non-linear effects change this value
such that $\alpha$ can take any value in the interval $(1,3)$
\cite{Rachen}. It is shown in \cite{Lipari:2007su} that for the
case of pion decay chain and assuming $\alpha=2$, the initial
flavor ratio is $w_e:w_\mu:w_\tau=1:1.85:0$ (the difference with
$1:2:0$ comes from the wrong polarization states of $\mu^\pm$ in
the decay of $\pi^\pm$). Also the authors of \cite{Pakvasa:2007dc}
show that inclusion of other secondary particles (such as $K^\pm$)
has a very little effect on this value. The $\mu^\pm$ generated in
the decays of the secondary particles can substantially lose their
energy before decay. In this case the neutrinos generated in the
decay of muons do not contribute to the flux of neutrinos with 100
GeV $<E_\nu< 100$ TeV and the flavor ratio becomes $0:1:0$.

In the calculation of $R$ for different scenarios of neutrino
production at source and propagation between source and Earth, the
uncertainties in the input parameters should be considered.
Uncertainties of the input parameters in the calculation of the
$\mu$-track and shower-like event rates induce uncertainties in
the value of $R$. Here we summarize these sources of
uncertainties:

\begin{description}
    \item[Mixing Parameters] Flavor content of the neutrino beam
    arriving at Earth depends on the mixing angles
    $(\theta_{12},\theta_{23},\theta_{13})$ and Dirac CP-violating
    phase $\delta$ through the $|U_{\alpha j}|^2|U_{\beta j}|^2$
    factors in Eq.~(\ref{flavor}). An update on the values of these
    parameters and uncertainties in each of them can be
    found in \cite{Schwetz:2008er}, which are also listed in Table~\ref{para}.
    \item[Spectral Index] As it is mentioned after Eq.~(\ref{powerlaw}),
    the spectral index $\alpha$ can take any value in the interval
    $(1,3)$. It is shown in \cite{Beacom:2003nh} that IceCube can measure the
    spectral index $\alpha$ with 10~\% precision (assuming $E_\nu^2dF_\nu/dE_\nu=
    0.25$~GeV~cm$^{-2}$~sr$^{-1}$~yr$^{-1}$ and after one year of
    data-taking).
    \item[$\bar{\nu}_\alpha/\nu_\alpha$ Ratio] Let us define
    $\lambda_\beta\equiv \mathcal{N}_{\bar{\nu}_\beta}/\mathcal{N}_{\nu_\beta}$ for
    $\beta=e,\mu$. It is obvious that $\lambda_\mu=1$ in the pion
    decay chain, but the value of $\lambda_e$ depends on the ratio
    of $\pi^+/\pi^-$ production and can take any value in the
    interval $(0,1)$. To best of our knowledge, in the energy range we are
    interested here (100 GeV $<E_\nu<100$ TeV), there is not any
    proposed or established method to measure $\lambda_e$.
    \item[Neutrino-Nucleon Cross Section] The rate of the
    $\mu$-track and shower-like events depends on the CC and NC
    neutrino-nucleon cross sections $\sigma^{CC}_{\nu N}$ and
    $\sigma^{NC}_{\nu N}$. The current uncertainty in these cross
    sections is $\sim 3~\%$. However, the uncertainties in $\sigma^{CC}_{\nu N}$ and
    $\sigma^{NC}_{\nu N}$ have a very small effect on $R$ because of the
    cancelation between numerator and denominator of
    Eq.~(\ref{ratioR}).
    \end{description}

\begin{table}[t]

\vspace{0.5 cm}
\begin{tabular}{|c|c|c|c|}
  \hline
  Parameter & Best-fit & Current Allowed Range & Future Uncertainty  \\
  \hline
  $\sin^2\theta_{12}$ & $0.304$ & $0.25-0.37$ (3$\sigma$ C.L.) & 6~\% (Ref. \cite{Ayres:2004js}) \\
  \hline
  $\sin^2\theta_{23}$ & $0.50$ & $0.36-0.67$ (3$\sigma$ C.L.) & 6~\% (Ref. \cite{:2008ee}) \\
  \hline
  $\sin^2\theta_{13}$ & $0.01$ & $\leq0.056\quad\,\;$ (3$\sigma$ C.L.) & 5~\% (Ref. \cite{Ardellier:2006mn}) \\
  \hline
  $\delta$ & $-$ & $[0,2\pi)$ & $[0,2\pi)$ \\
  \hline
  $\alpha$ & $-$ & 10~\% & 10~\% \\
  \hline
  $\lambda_e\equiv\mathcal{N}_{\bar{\nu}_e}/\mathcal{N}_{\nu_e}$ & $-$ & $[0,1]$ & $[0,1]$ \\
  \hline
\end{tabular}
\centering
  \caption{Relevant parameters in the calculation of $R$ with
  their uncertainties. The current uncertainty column represents
  the $3\sigma$ uncertainty interval for mixing
angles \cite{Schwetz:2008er}. The future uncertainty column shows
the precisions that can be achieved in the forthcoming neutrino
oscillation experiments, as described in the corresponding
references. For $\theta_{13}$ it is assumed that its value is
close to the current upper limit:
$\sin^2\theta_{13}=0.03$.}\label{para} \vspace{1 cm}
\end{table}

In Table~\ref{para} we have listed the relevant parameters in the
calculation of $R$ and their uncertainty intervals. The third and
forth columns respectively correspond to the current and future
uncertainties in these parameters. In addition to the
uncertainties in the input parameters listed above, measurement of
$R$ is also done with limited precision and has an uncertainty. It
is shown in \cite{Beacom:2003nh} that by assuming that the flux of
neutrinos is
$E_\nu^2dF_\nu/dE_\nu=0.25$~GeV~cm$^{-2}$~sr$^{-1}$~yr$^{-1}$, the
ratio $R$ can be measured with $\sim~7$~\% precision after a
couple of years of data-taking.

Now, by taking into account the uncertainties mentioned in this
section, the question is to what extent it is possible to measure
the deviations of the flavor composition
$F_{\nu_e}:F_{\nu_\mu}:F_{\nu_\tau}$ from its value in the absence
of pseudo-Dirac sterile neutrinos. In the next section we will
discuss the prospect of neutrino telescopes to measure these
deviations.

\section{Results and Discussion \label{result}}

In this section we consider two sources with different initial
flavor compositions: the pion and stopped-muon sources with flavor
compositions $1:1.85:0$ and $0:1:0$, respectively. The value of
$R$ in Eq.~(\ref{ratioR}) in the absence of almost degenerate
sterile neutrinos and assuming the best-fit values for the mixing
angles, $\delta=0$ and $\lambda_e=1$, will be denoted by
$\bar{R}_{\pi}$ and $\bar{R}_{\mu}$, for pion and stopped-muon
sources respectively. We assume the power-law spectrum
Eq.~(\ref{powerlaw}) for the neutrino production, with the
spectral index $\alpha=2$. At the end, we investigate the
robustness of the results with respect to the deviation of initial
flavor ratios from the assumed values $1:1.85:0$ and $0:1:0$.

\begin{figure}[p]
  \begin{center}
 \centerline{\includegraphics[bb=310 70 570
 500,keepaspectratio=true,clip=true,angle=-90,scale=0.6]{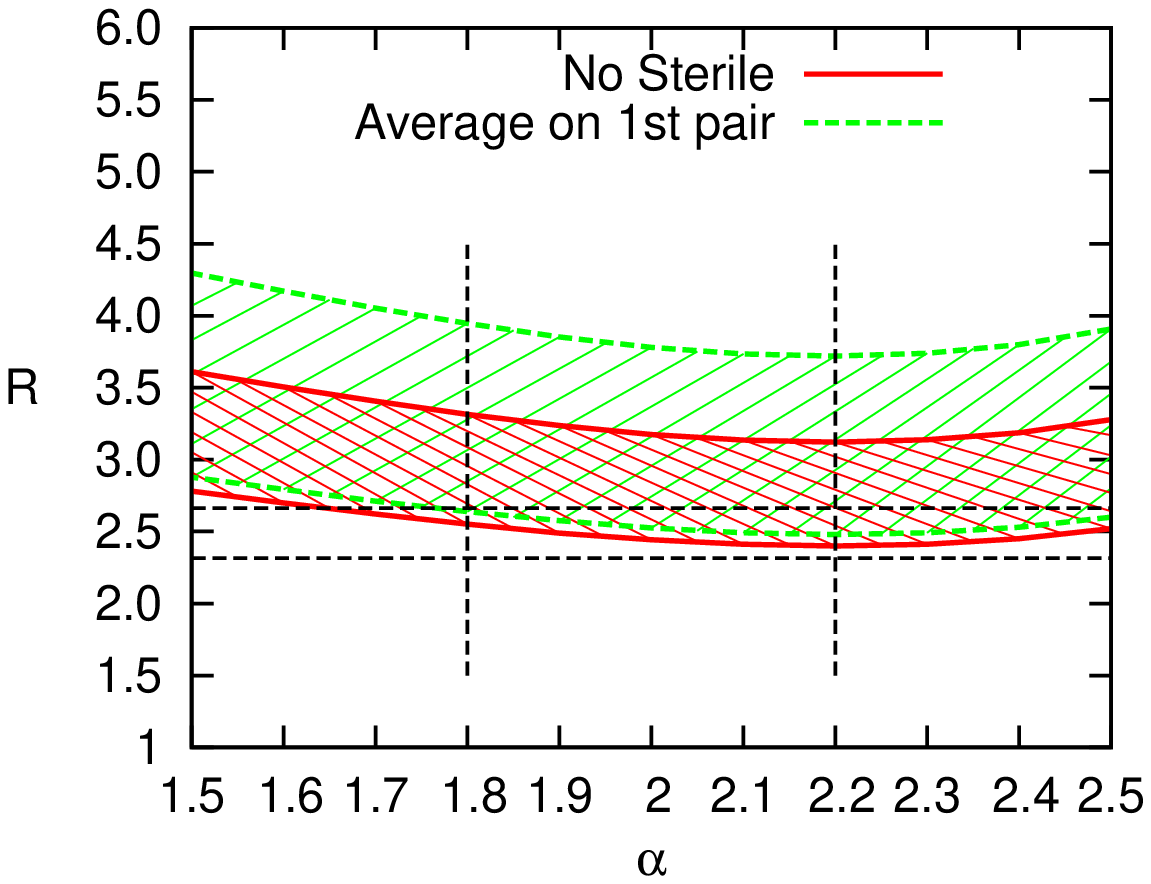}\includegraphics[bb=310 70 570
 450,keepaspectratio=true,clip=true,angle=-90,scale=0.6]{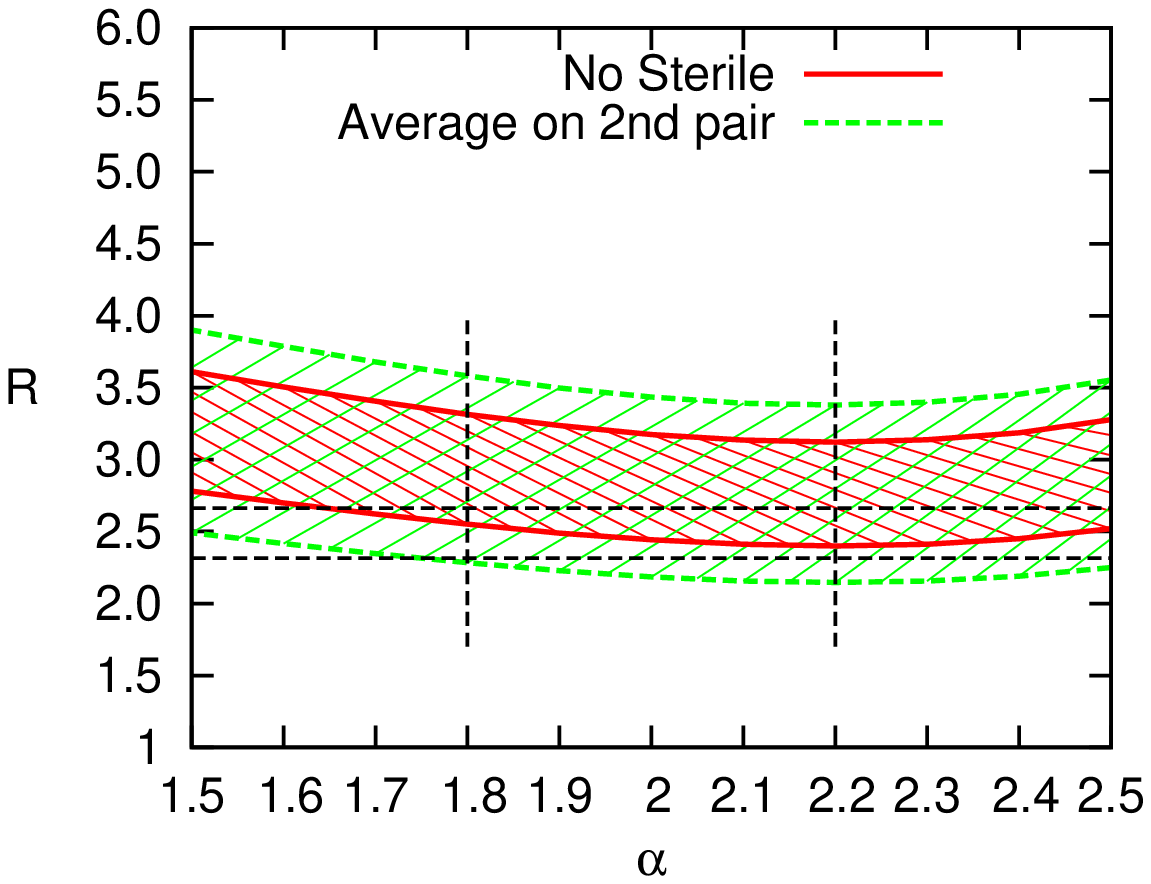}}
 \centerline{\hspace{0.45cm}(a)\hspace{8.45cm}(b)}
 \centerline{\vspace{-2.0cm}}
 \centerline{\includegraphics[bb=260 70 570
 500,keepaspectratio=true,clip=true,angle=-90,scale=0.6]{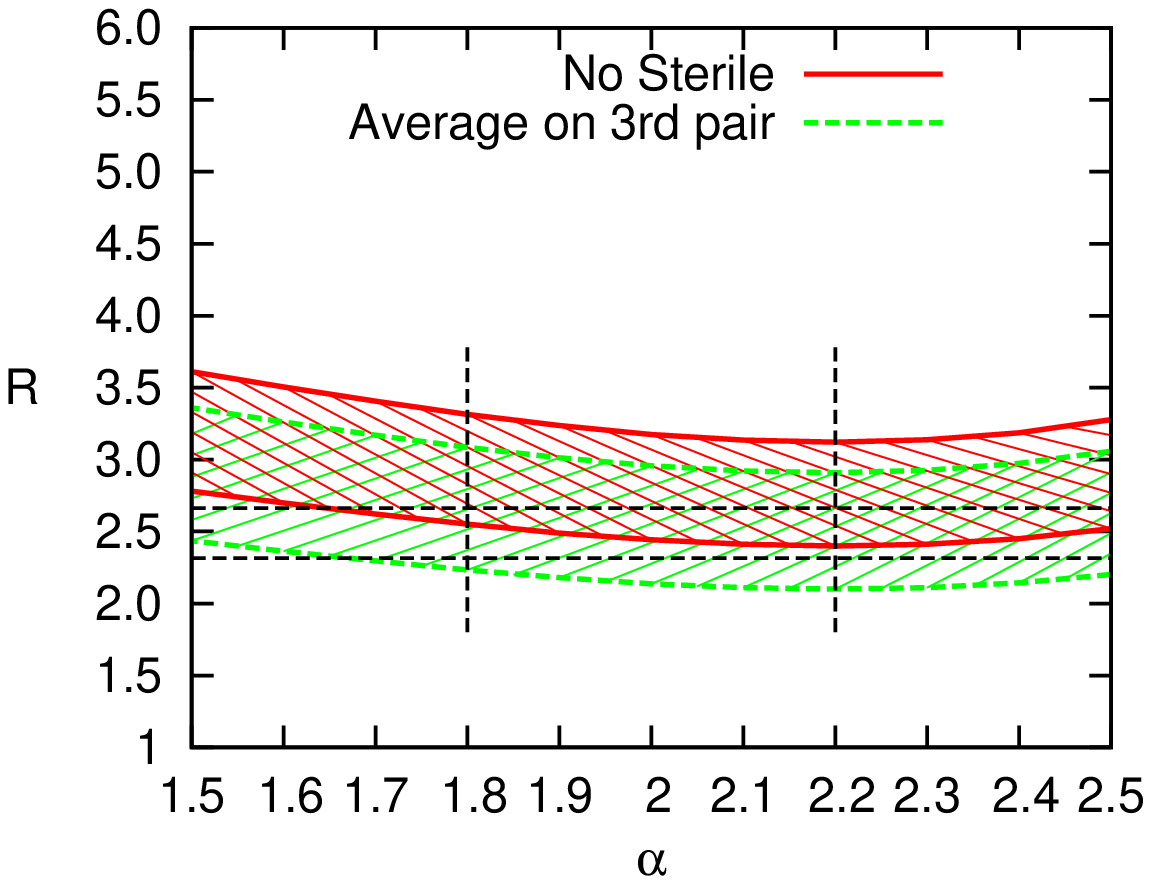}\includegraphics[bb=260 70 570
 450,keepaspectratio=true,clip=true,angle=-90,scale=0.6]{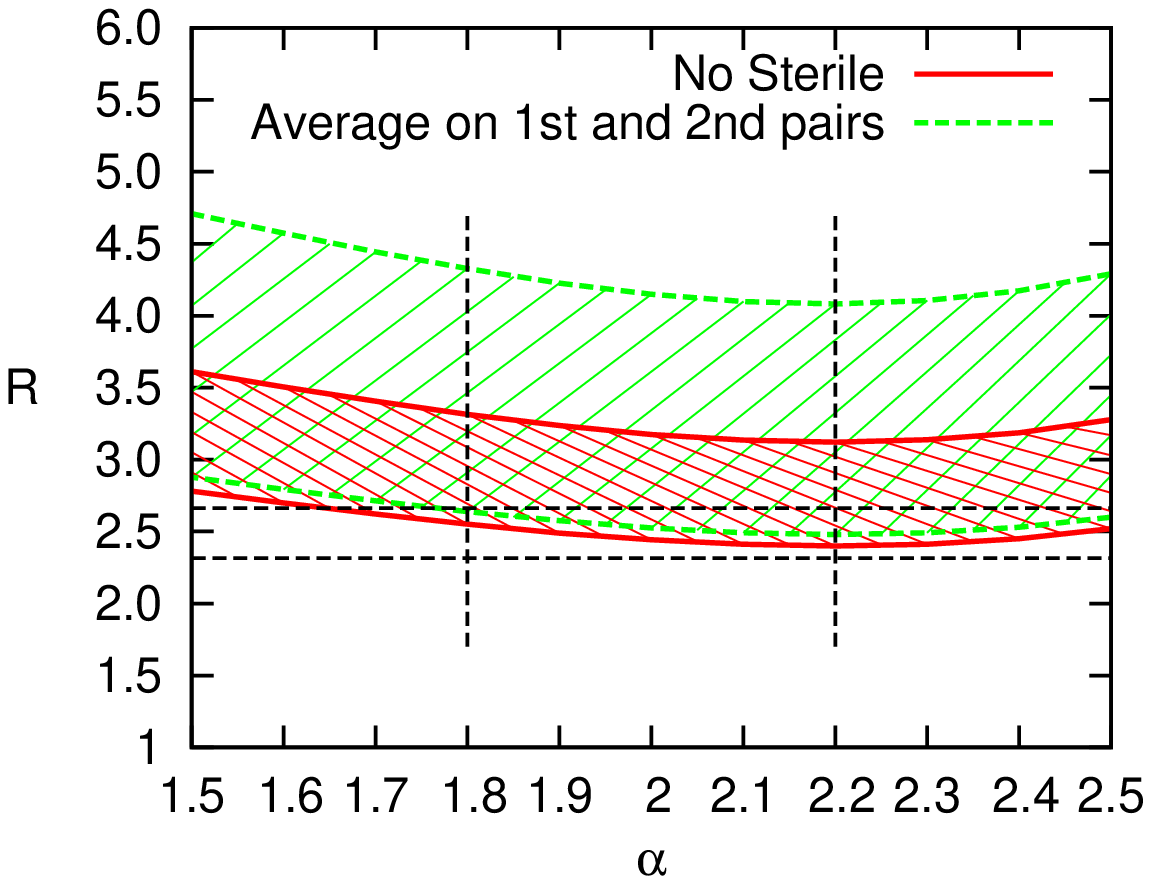}}
 \centerline{\hspace{0.45cm}(c)\hspace{8.45cm}(d)}
 \centerline{\vspace{-2.0cm}}
 \centerline{\includegraphics[bb=260 70 570
 500,keepaspectratio=true,clip=true,angle=-90,scale=0.6]{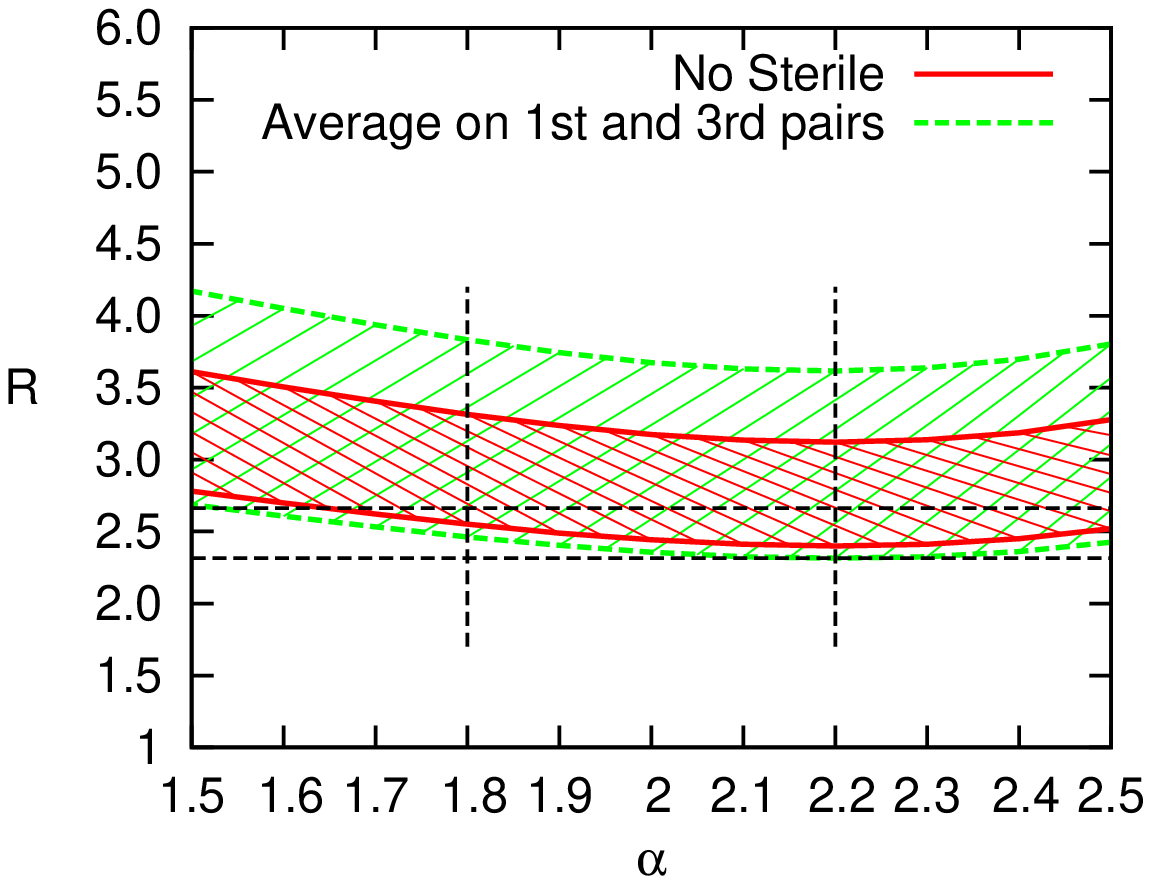}\includegraphics[bb=260 70 570
 450,keepaspectratio=true,clip=true,angle=-90,scale=0.6]{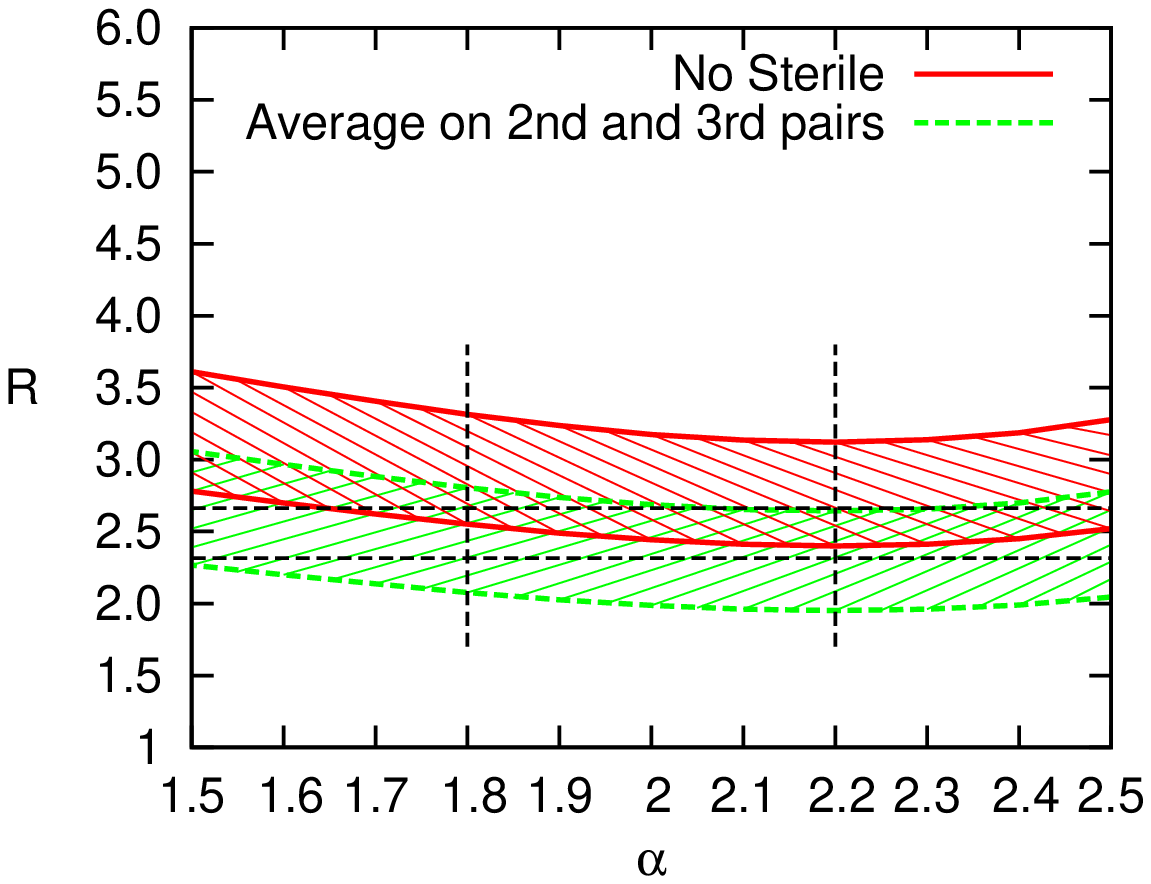}}
 \centerline{\hspace{0.45cm}(e)\hspace{8.45cm}(f)}
 \centerline{\vspace{-1.7cm}}
 \end{center}
 \caption{{\small The dependence of $R$, ratio of $\mu$-tracks to shower-like
 events, on $\alpha$ for the pion source with
 the initial flavor ratio $1:1.85:0$ and power-law spectrum with the
 spectral index $\alpha=2$. Assuming the
 best-fit values for the mixing angles, $\delta=0$ and
 $\lambda_e=1$, the value of $R$ is $\bar{R}_\pi=2.50$. In each figure the red curves
 represent the case with no sterile neutrino and the green dashed-curves
 correspond to the cases with average on pairs mentioned in the
 legends. The hatched areas show the values that $R$ can take when
 the input parameters vary in the current uncertainty intervals in
 Table~\ref{para}. The two vertical and horizontal dashed-lines
 show the 10~\% and 7~\% precisions in the measurements of
 $\alpha$ and $R$, respectively.
 } }
  \label{120}
\end{figure}

\begin{figure}[p]
  \begin{center}
 \centerline{\includegraphics[bb=310 70 570
 500,keepaspectratio=true,clip=true,angle=-90,scale=0.6]{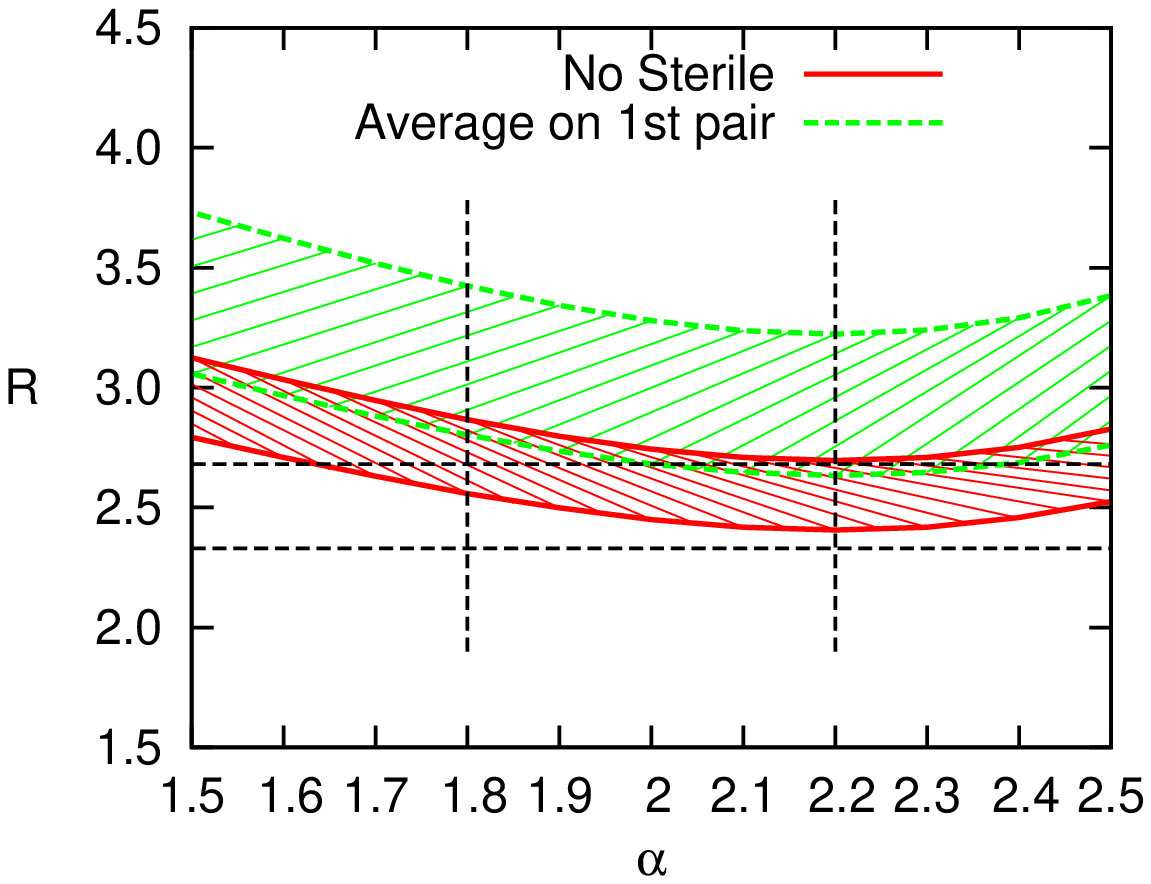}\includegraphics[bb=310 70 570
 450,keepaspectratio=true,clip=true,angle=-90,scale=0.6]{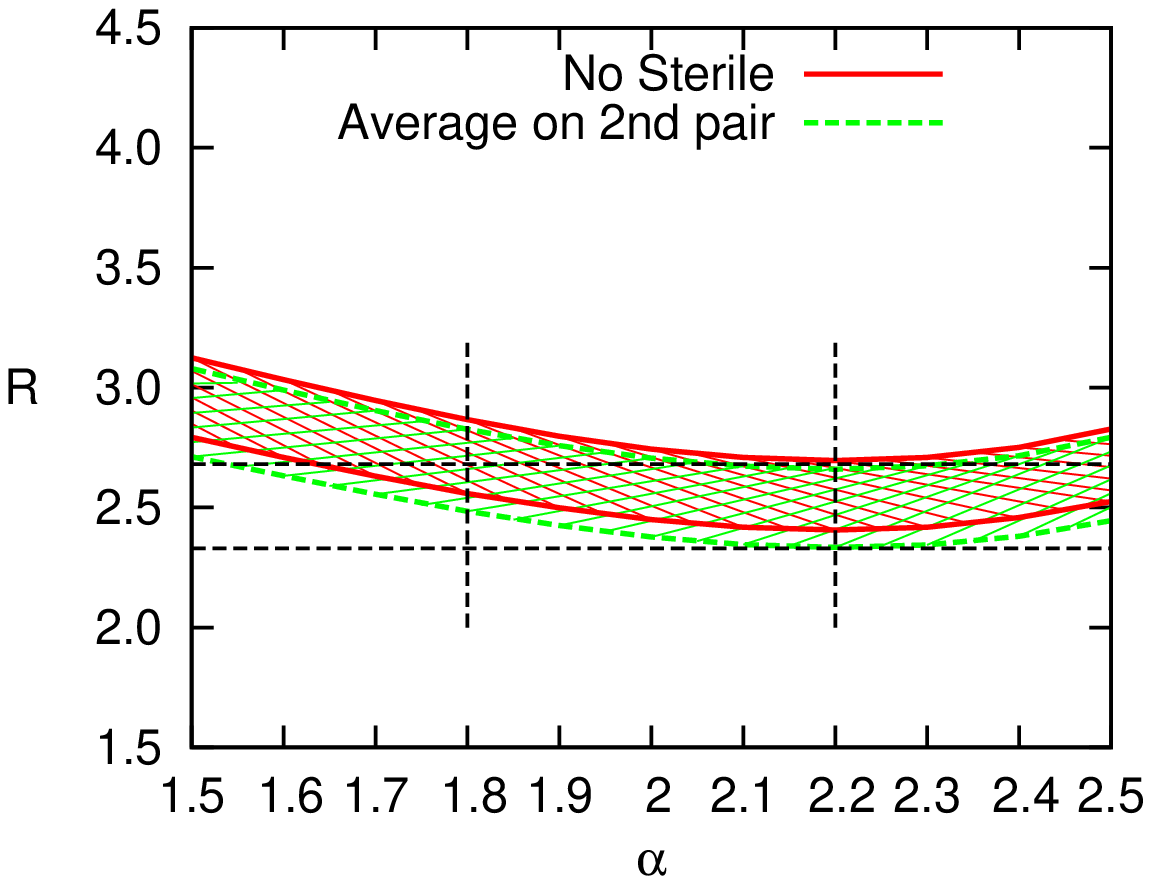}}
 \centerline{\hspace{0.45cm}(a)\hspace{8.45cm}(b)}
 \centerline{\vspace{-2.0cm}}
 \centerline{\includegraphics[bb=260 70 570
 500,keepaspectratio=true,clip=true,angle=-90,scale=0.6]{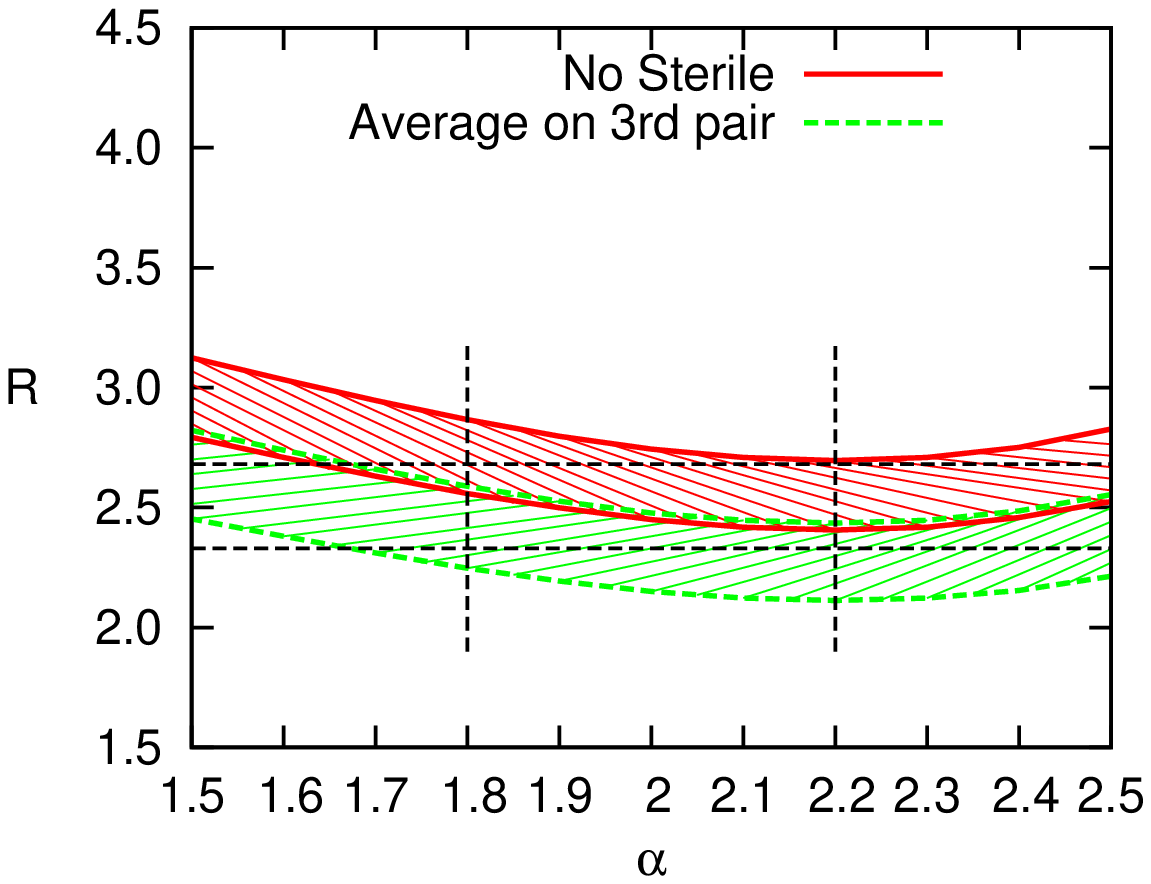}\includegraphics[bb=260 70 570
 450,keepaspectratio=true,clip=true,angle=-90,scale=0.6]{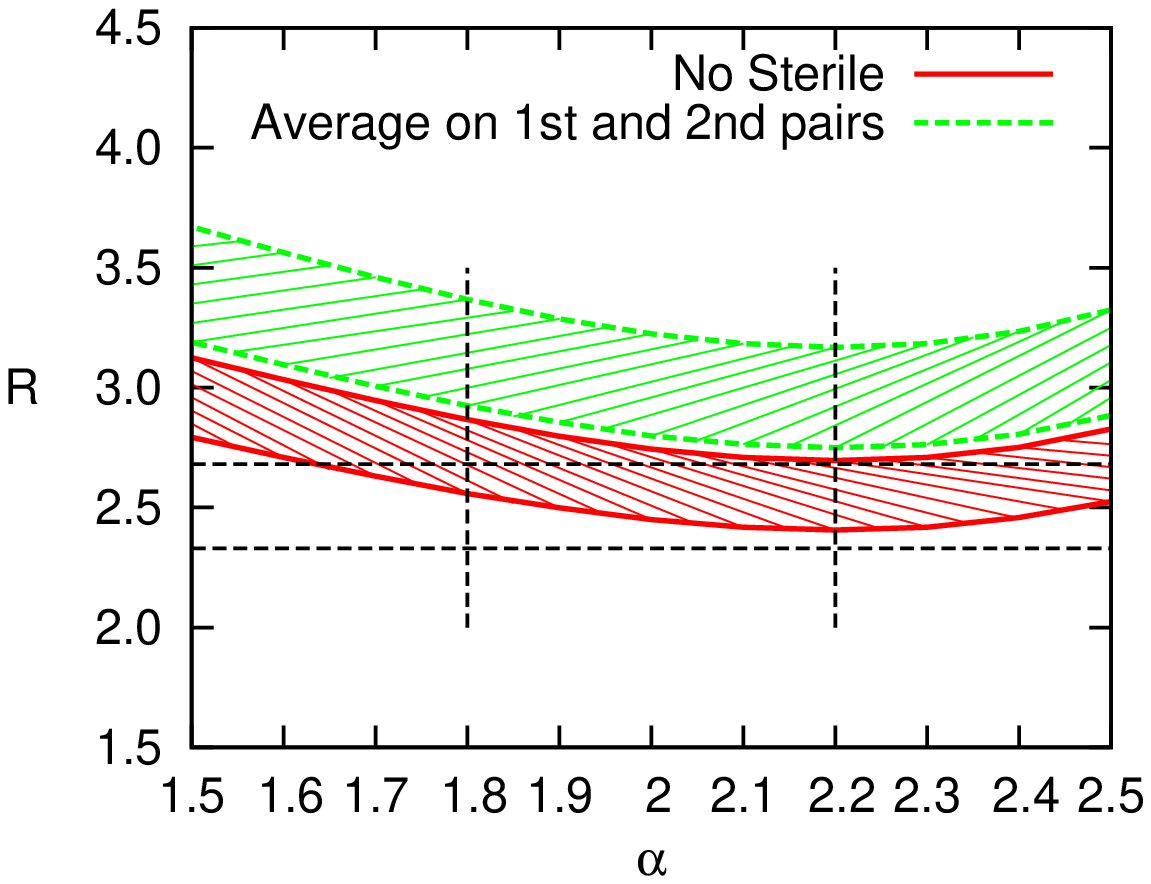}}
 \centerline{\hspace{0.45cm}(c)\hspace{8.45cm}(d)}
 \centerline{\vspace{-2.0cm}}
 \centerline{\includegraphics[bb=260 70 570
 500,keepaspectratio=true,clip=true,angle=-90,scale=0.6]{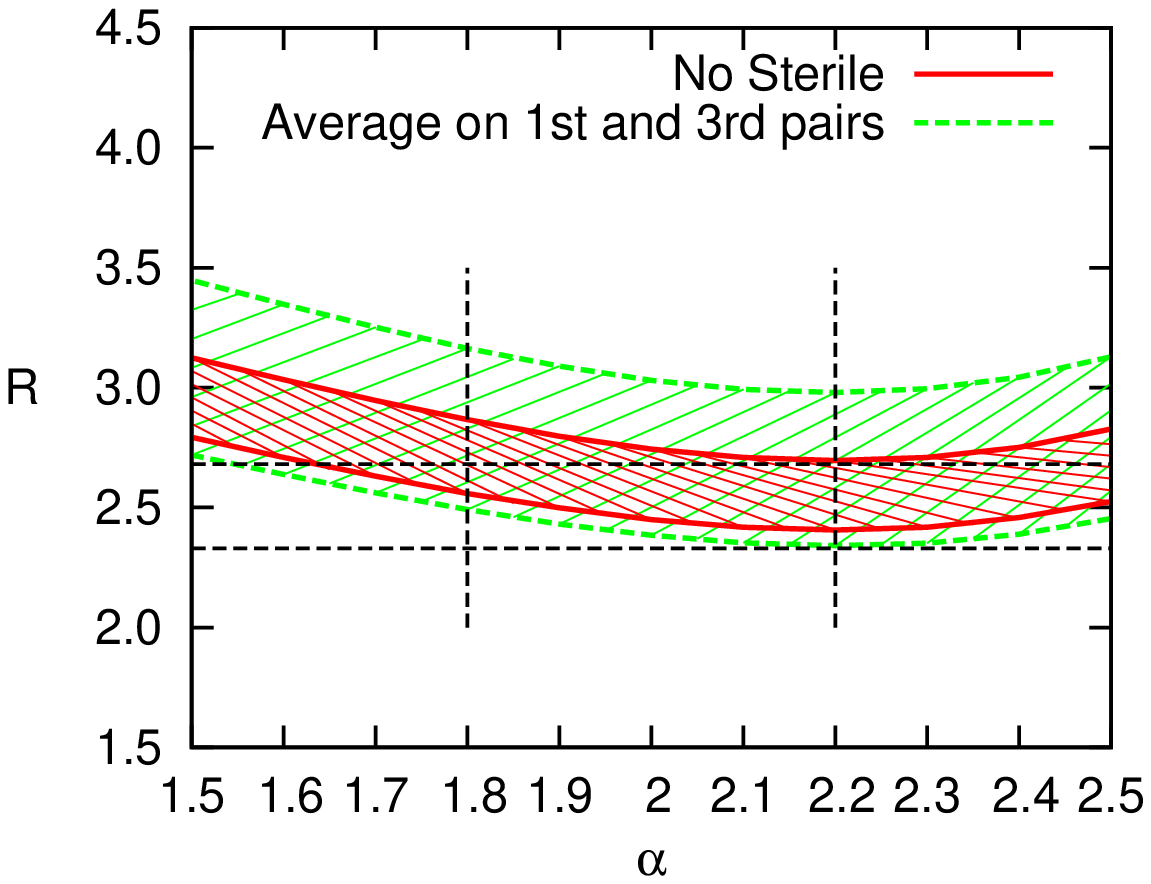}\includegraphics[bb=260 70 570
 450,keepaspectratio=true,clip=true,angle=-90,scale=0.6]{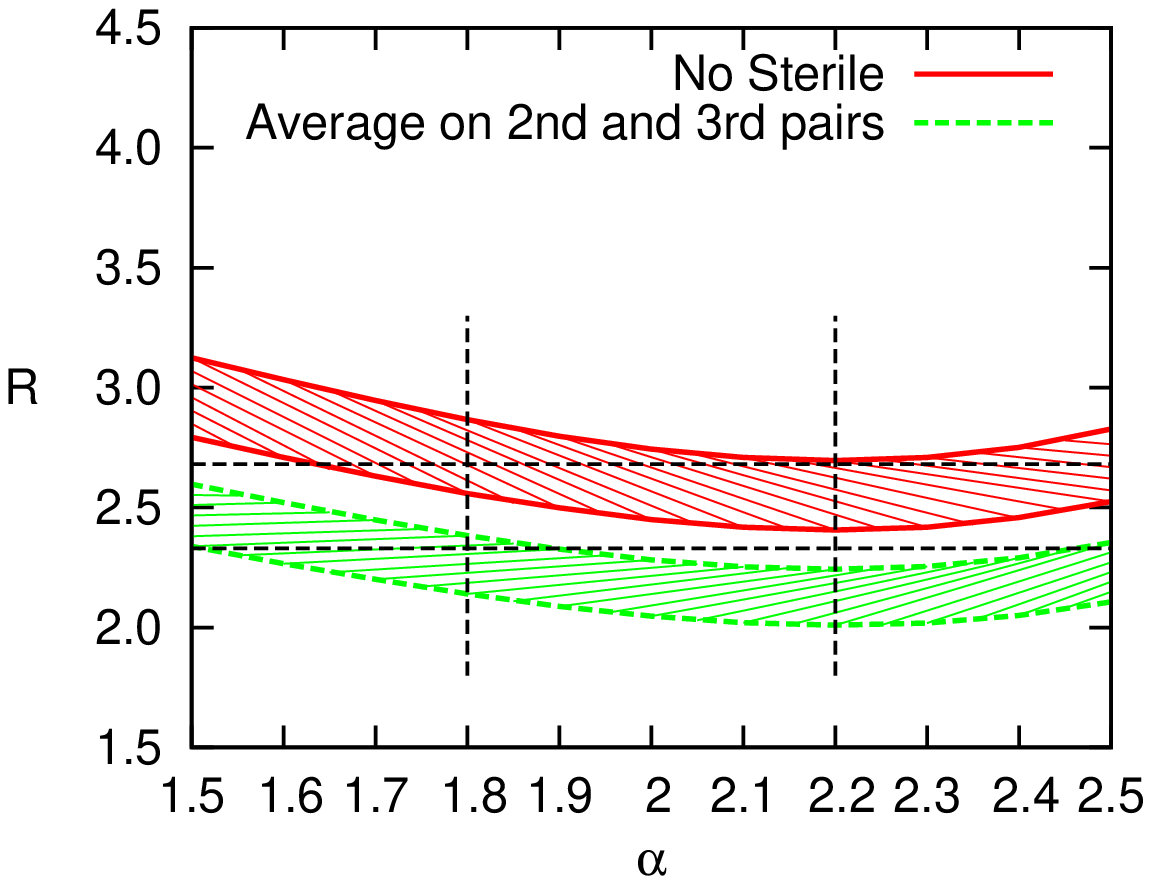}}
 \centerline{\hspace{0.45cm}(e)\hspace{8.45cm}(f)}
 \centerline{\vspace{-1.7cm}}
 \end{center}
 \caption{{\small The same as Fig.~(\ref{120}) except that the
 input parameters have been varied in the future uncertainty
 intervals of Table~\ref{para}. Particularly, we assumed 5~\%
 uncertainty interval for the 13-mixing angle with the central
 value near the present upper bound: $\sin^2\theta_{13}=0.03$.
 } }
  \label{120new}
\end{figure}

Fig.~(\ref{120}) shows $R$ versus $\alpha$ for pion source, where
$\bar{R}_\pi=2.50$. The hatched area between the two red curves
corresponds to the values that $R$ can take if the input
parameters vary in the uncertainty intervals. The inputs have been
varied in the current uncertainty intervals shown in the third
column of Table~\ref{para}. The two horizontal and vertical pairs
of dashed-lines show the 7~\%  and 10~\% precision intervals in
the measurement of $R$ and $\alpha$, respectively. The rectangle
created from the intersection of these dashed-lines corresponds to
the region of parameter space $(R,\alpha)$ where can be limited by
the measurements in neutrino telescopes. Thus, the points inside
the rectangle represent the values of $R$ for the case of no
sterile neutrino and consistent with 7~\% (10~\%) precision in the
measurement of $R$ ($\alpha$). The green dashed-curves in
Fig.~(\ref{120}) show the values of $R$ in the presence of sterile
neutrinos. For example, in Fig.~(\ref{120}-a) the hatched region
between the green dashed-curves corresponds to the case when the
sterile neutrino mass is almost degenerate with the mass of the
active neutrino $\nu_{1L}$ such that the oscillatory term
depending on $\Delta m^2_1$ in Eq.~(\ref{probability}) can be
averaged to $1/2$ and the flavor conversion probabilities become
\begin{equation} P_{\alpha\beta}=\frac{1}{2}|U_{\alpha
1}|^2|U_{\beta 1}|^2+ |U_{\alpha 2}|^2|U_{\beta 2}|^2+ |U_{\alpha
3}|^2|U_{\beta 3}|^2.\end{equation} As can be seen, in all six
parts of the Fig.~(\ref{120}) the two hatched areas overlap. The
overlapping of the hatched areas inside the dashed-line rectangles
means that the presence of the sterile neutrinos cannot be ruled
out, even if the measurement of $R$ gives a value inside this
rectangles. However, if the measurement of $R$ gives a value much
different than $\bar{R}_\pi$, the existence of sterile neutrinos
is favored. For example, the value $R=2.1$ is not possible in the
no sterile case and this value is favored by the scenarios
depicted in Fig.~(\ref{120}-b,c,f).

Fig.~(\ref{120new}) is the same as Fig.~(\ref{120}) with the
exception that the input parameters have been varied in the future
uncertainty intervals shown in the forth column of
Table~\ref{para}. We assumed (6\%,~6\%,~5\%) uncertainties for
($\sin^2\theta_{12},\sin^2\theta_{23},\sin^2\theta_{13}$), which
can be achieved in the forthcoming neutrino oscillation
experiments \cite{Ayres:2004js,:2008ee,Ardellier:2006mn}. Also, we
assumed a large value for the 13-mixing angle
$\sin^2\theta_{13}=0.03$ which is near its present upper bound.
The uncertainties of $\delta$, $\lambda_e$, $\alpha$ and $R$ are
the same as in Fig.~(\ref{120}). As can be seen from
Fig.~(\ref{120new}), reducing the uncertainties of mixing angles
results in a separation between the two hatched areas where there
was an overlap in Fig.~(\ref{120}). For example, in
Fig.~(\ref{120new}-f), the regions corresponding to no sterile
neutrino case and the case with averaged second and third pairs
are completely separated. This separation means that if the
measurement of $R$ gives a value inside the rectangle in
Fig.~(\ref{120new}-f), the existence of sterile neutrinos (for the
case of average on second and third pairs) will be ruled out. On
the other hand, a value of $R$ outside the rectangle can be
interpreted as a signal for the existence of sterile neutrinos.
However, recognizing which case is consistent with the measurement
of $R$ should be done with care. To illuminate this point, let us
assume that IceCube has measured $R=3.0$. According to the
diagrams in Fig.~(\ref{120new}) this value of $R$ is consistent
with three cases: i) average on first pair; ii) average on first
and second pairs; and iii) average on first and third pairs.
Notice that with the current uncertainties of the mixing angles
(depicted in Fig.~(\ref{120})) it is not possible to conclude that
$R=3.0$ is a signal of sterile neutrinos.

\begin{figure}[p]
  \begin{center}
 \centerline{\includegraphics[bb=310 70 570
 500,keepaspectratio=true,clip=true,angle=-90,scale=0.6]{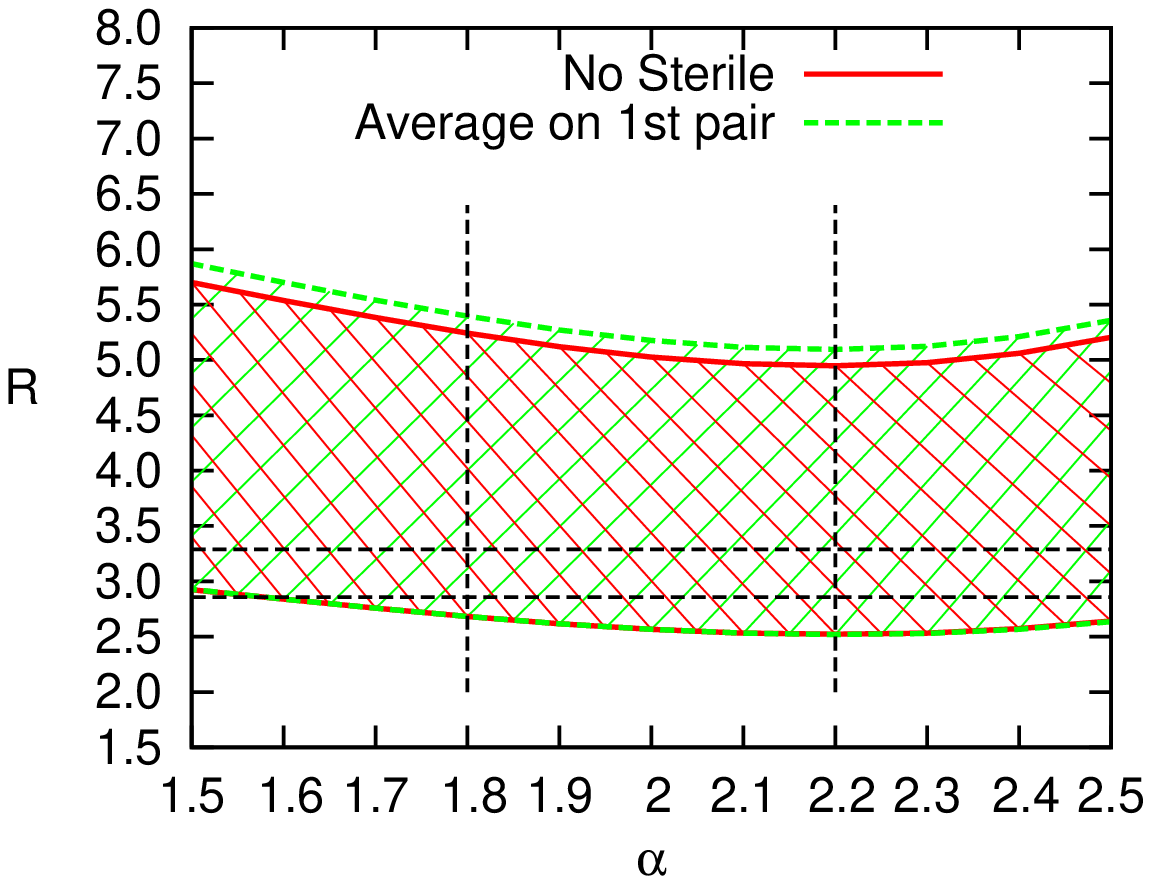}\includegraphics[bb=310 70 570
 450,keepaspectratio=true,clip=true,angle=-90,scale=0.6]{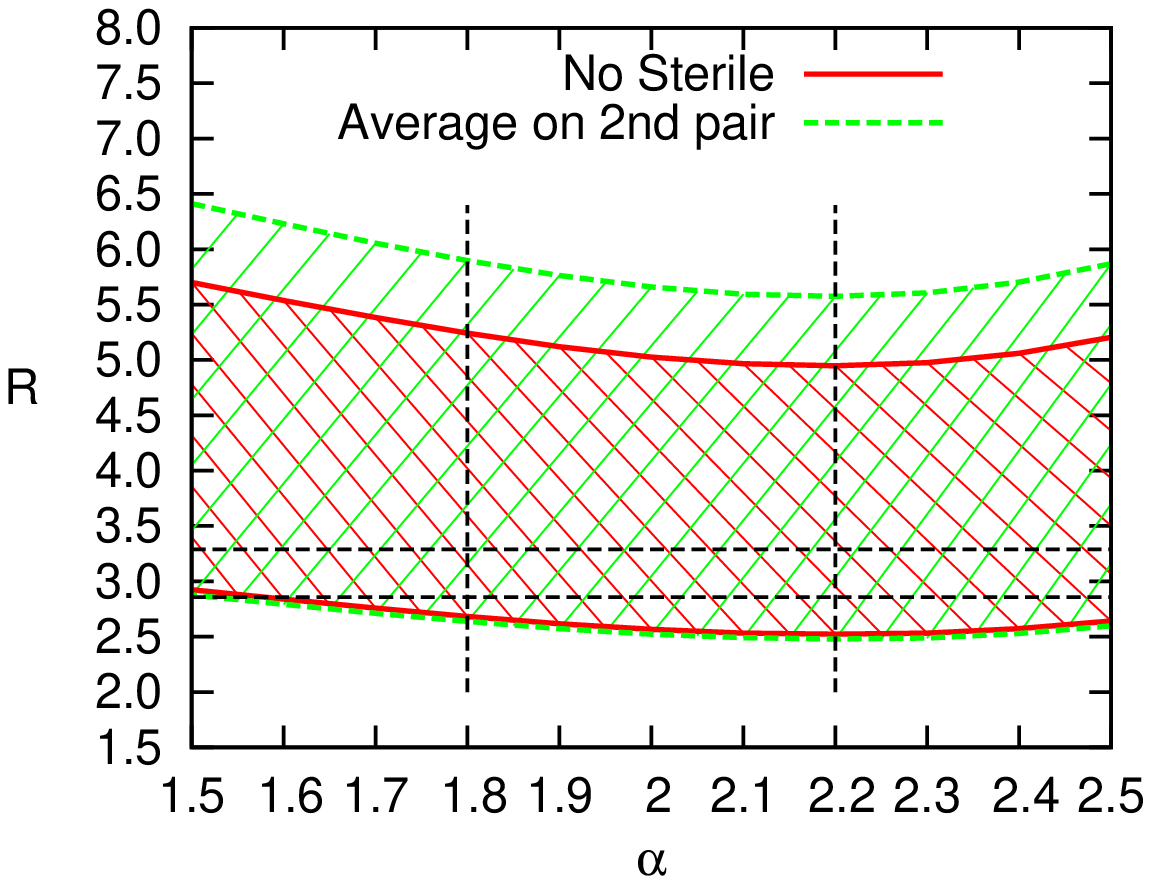}}
 \centerline{\hspace{0.45cm}(a)\hspace{8.45cm}(b)}
 \centerline{\vspace{-2.0cm}}
 \centerline{\includegraphics[bb=260 70 570
 500,keepaspectratio=true,clip=true,angle=-90,scale=0.6]{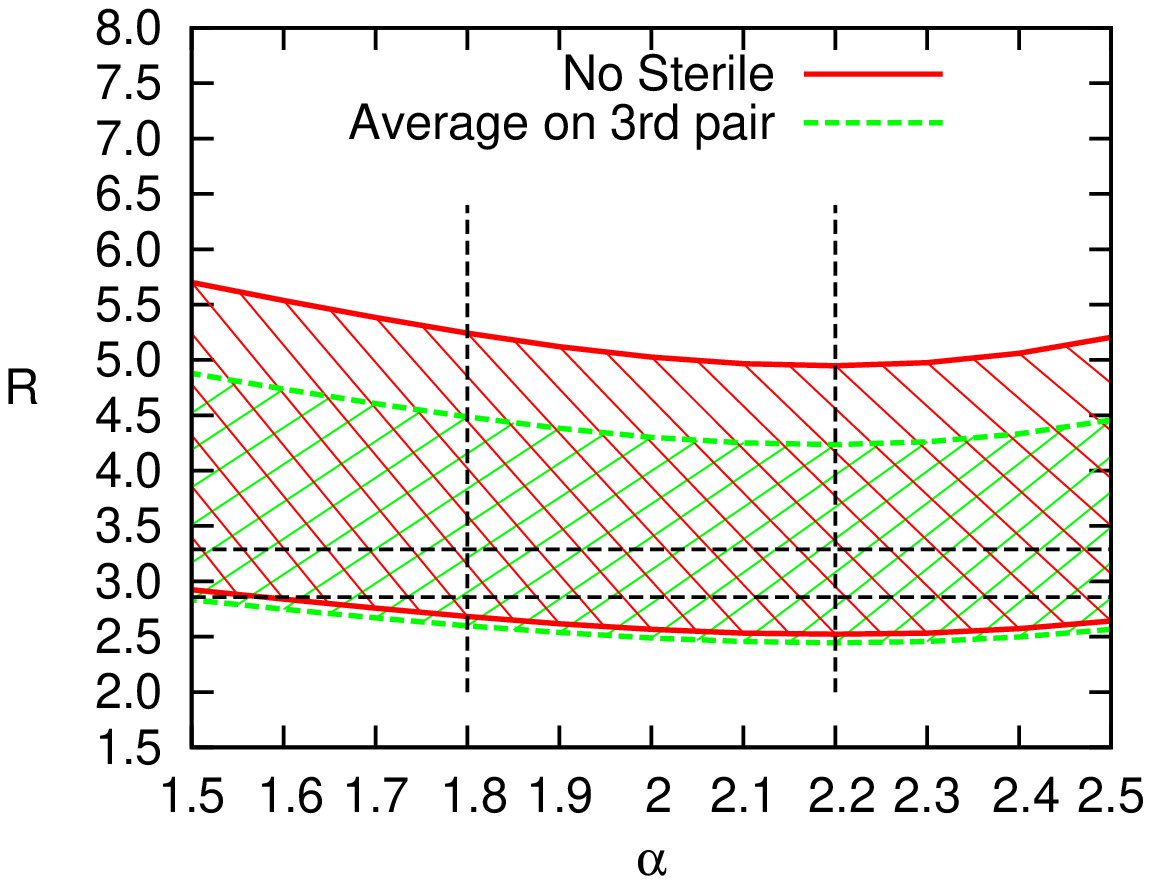}\includegraphics[bb=260 70 570
 450,keepaspectratio=true,clip=true,angle=-90,scale=0.6]{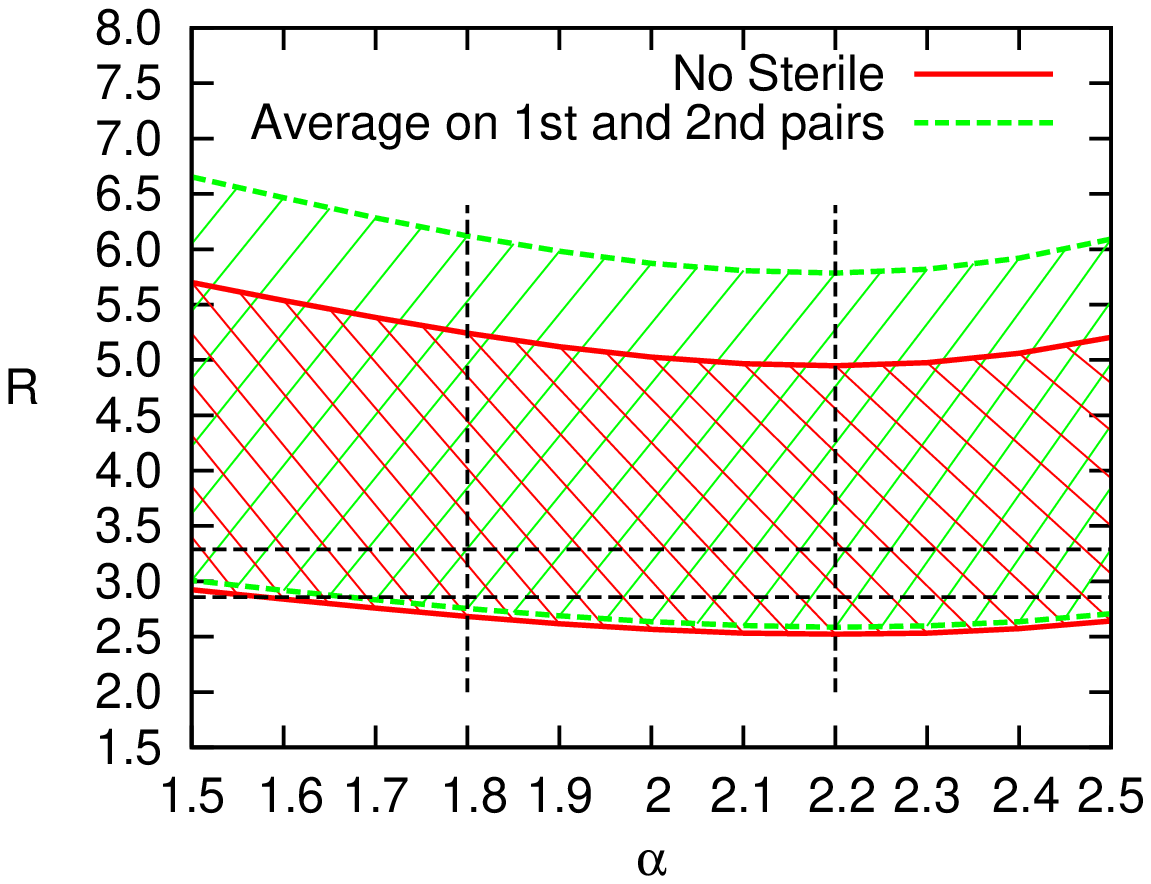}}
 \centerline{\hspace{0.45cm}(c)\hspace{8.45cm}(d)}
 \centerline{\vspace{-2.0cm}}
 \centerline{\includegraphics[bb=260 70 570
 500,keepaspectratio=true,clip=true,angle=-90,scale=0.6]{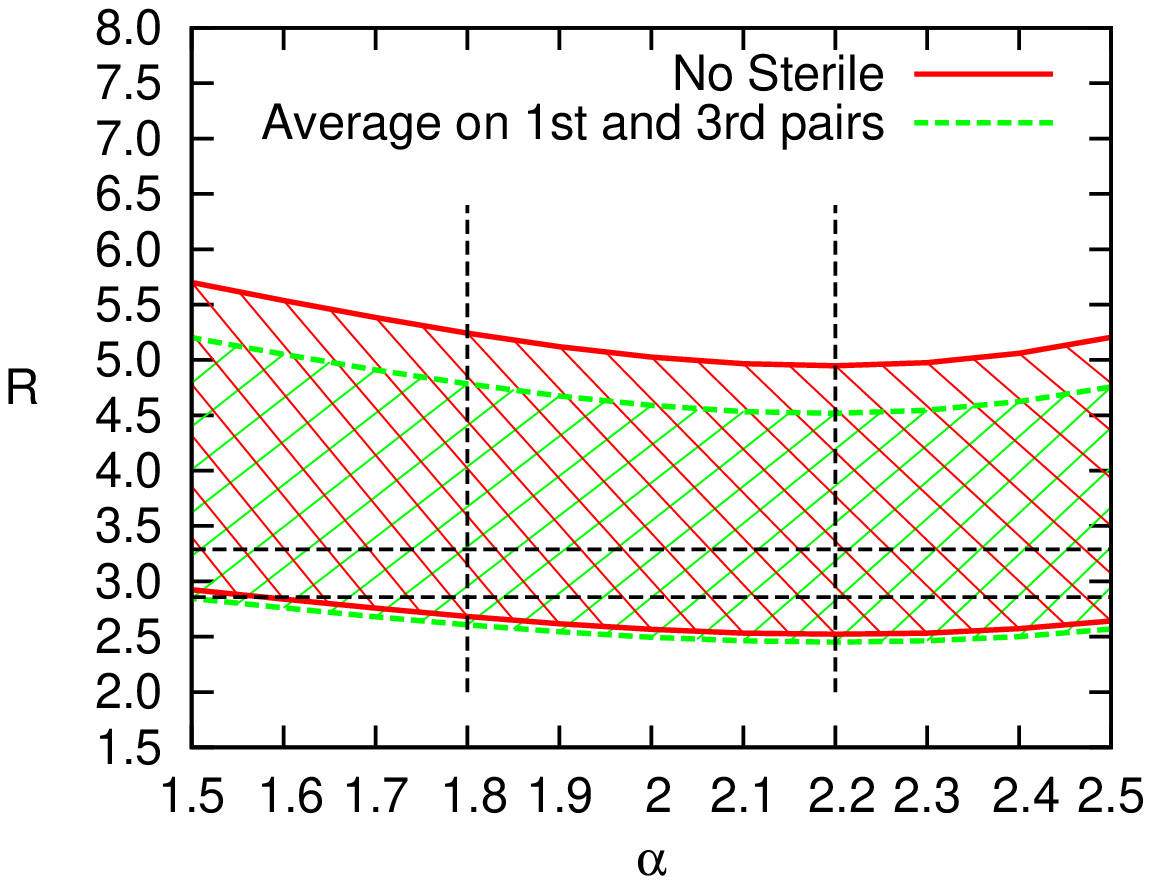}\includegraphics[bb=260 70 570
 450,keepaspectratio=true,clip=true,angle=-90,scale=0.6]{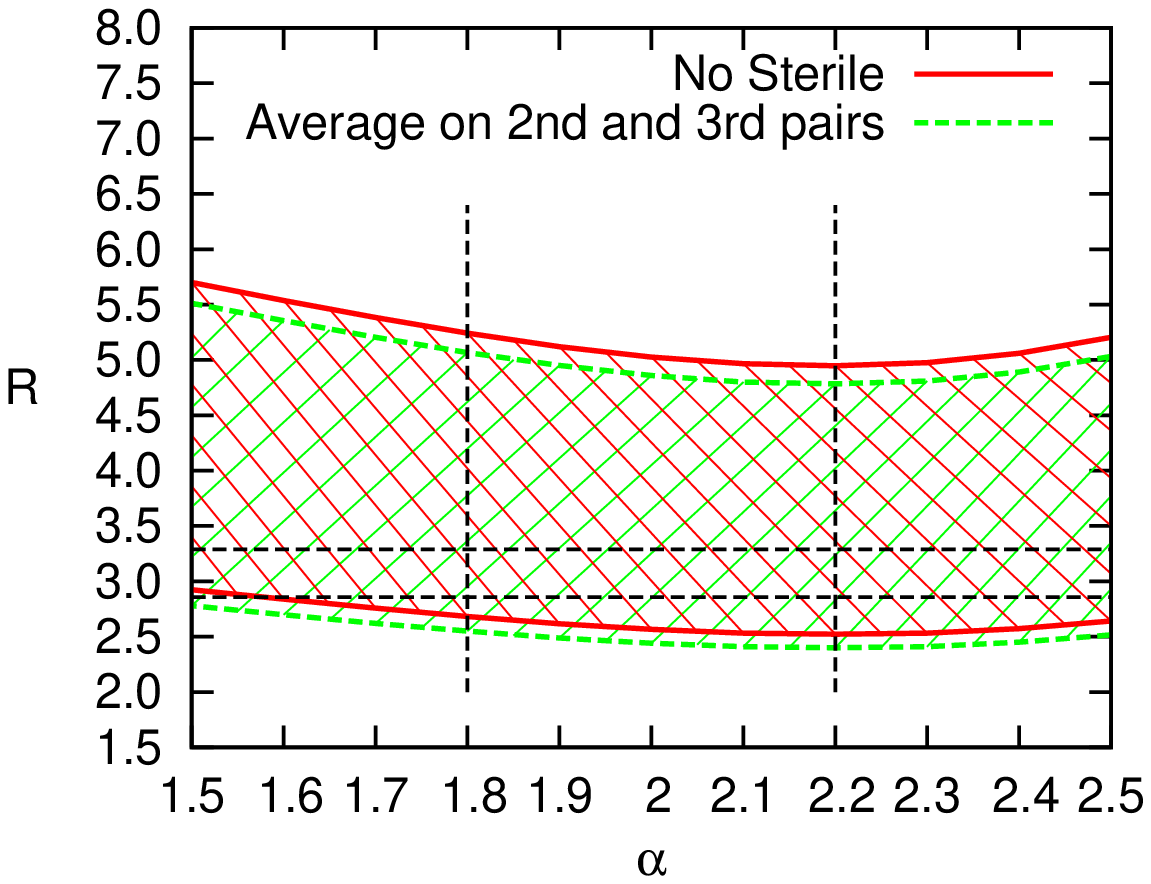}}
 \centerline{\hspace{0.45cm}(e)\hspace{8.45cm}(f)}
 \centerline{\vspace{-1.7cm}}
 \end{center}
 \caption{{\small The dependence of $R$, ratio of $\mu$-tracks to shower-like
 events, on $\alpha$ for the stopped-muon source with
 the initial flavor ratio $0:1:0$ and power-law spectrum with the
 spectral index $\alpha=2$. Assuming the
 best-fit values for the mixing angles and $\delta=0$,
 the value of $R$ is $\bar{R}_\mu=3.07$. In each figure the red curves
 represent the case with no sterile neutrino and the green dashed-curves
 correspond to the cases with average on pairs mentioned in the
 legends. The hatched areas show the values that $R$ can take when
 the input parameters vary in the current uncertainty intervals in
 Table~\ref{para}. The two vertical and horizontal dashed-lines
 show the 10~\% and 7~\% precisions in the measurements of
 $\alpha$ and $R$, respectively.
 } }
  \label{010}
\end{figure}

\begin{figure}[p]
  \begin{center}
 \centerline{\includegraphics[bb=310 70 570
 500,keepaspectratio=true,clip=true,angle=-90,scale=0.6]{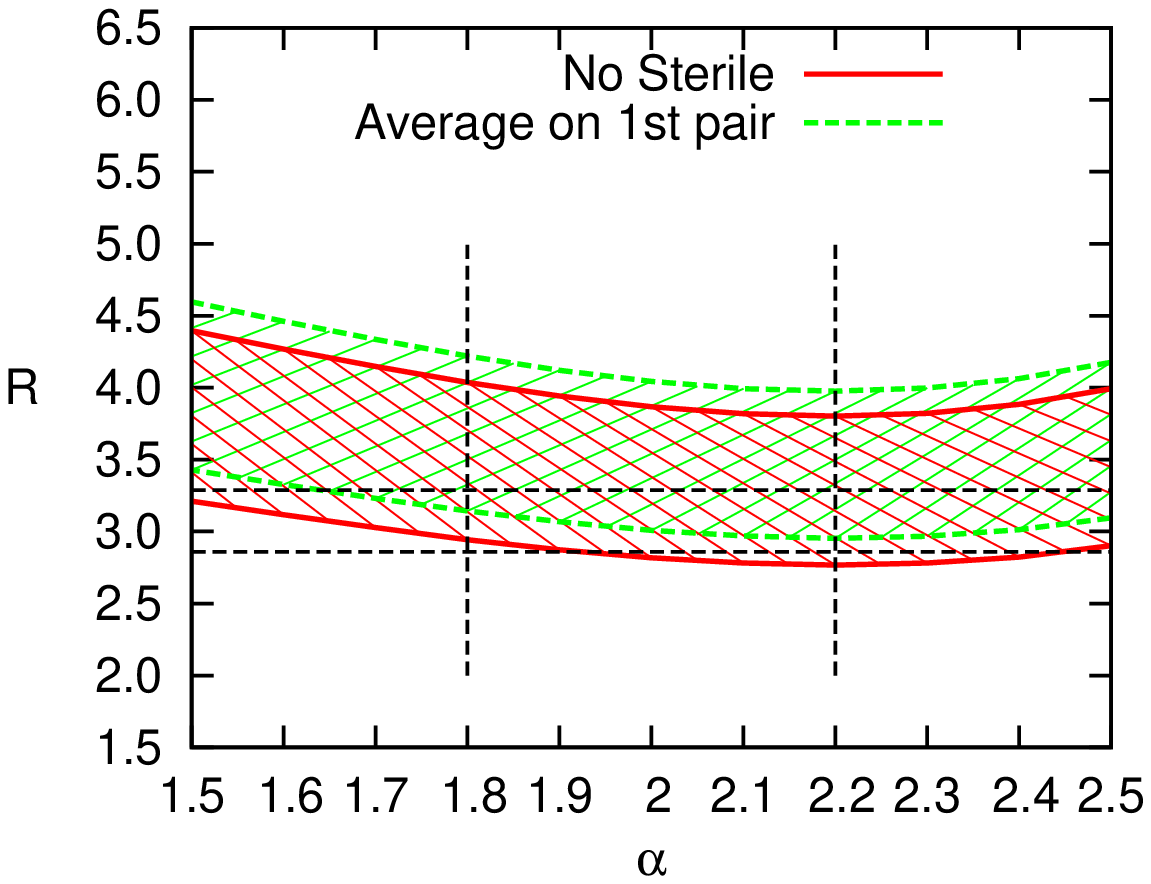}\includegraphics[bb=310 70 570
 450,keepaspectratio=true,clip=true,angle=-90,scale=0.6]{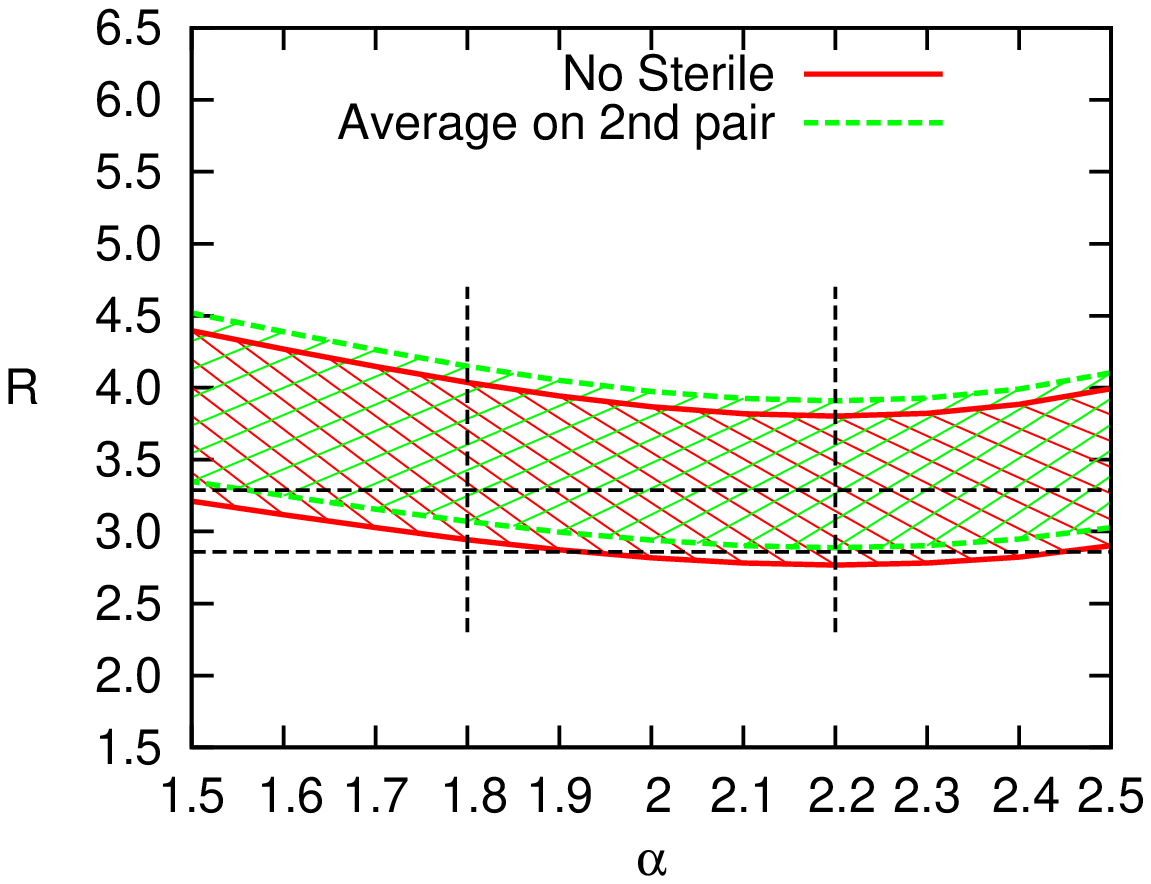}}
 \centerline{\hspace{0.45cm}(a)\hspace{8.45cm}(b)}
 \centerline{\vspace{-2.0cm}}
 \centerline{\includegraphics[bb=260 70 570
 500,keepaspectratio=true,clip=true,angle=-90,scale=0.6]{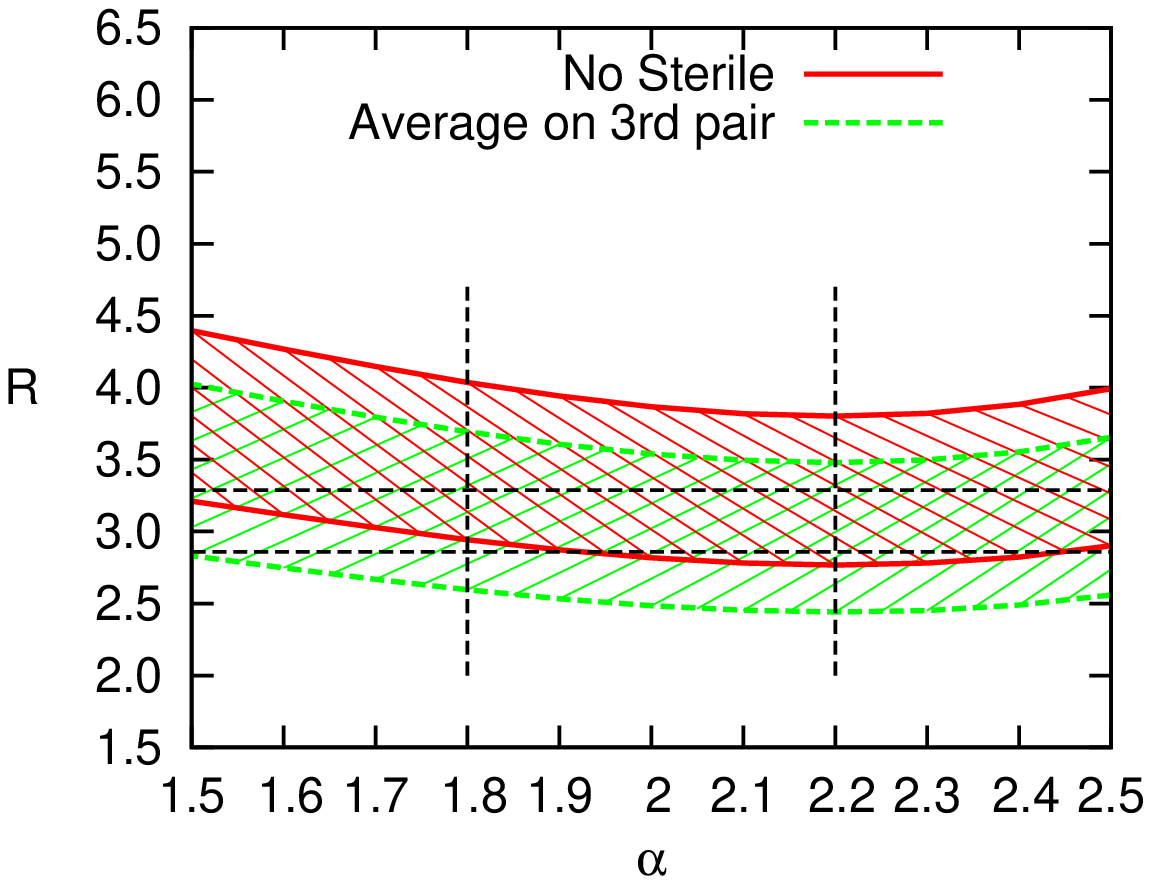}\includegraphics[bb=260 70 570
 450,keepaspectratio=true,clip=true,angle=-90,scale=0.6]{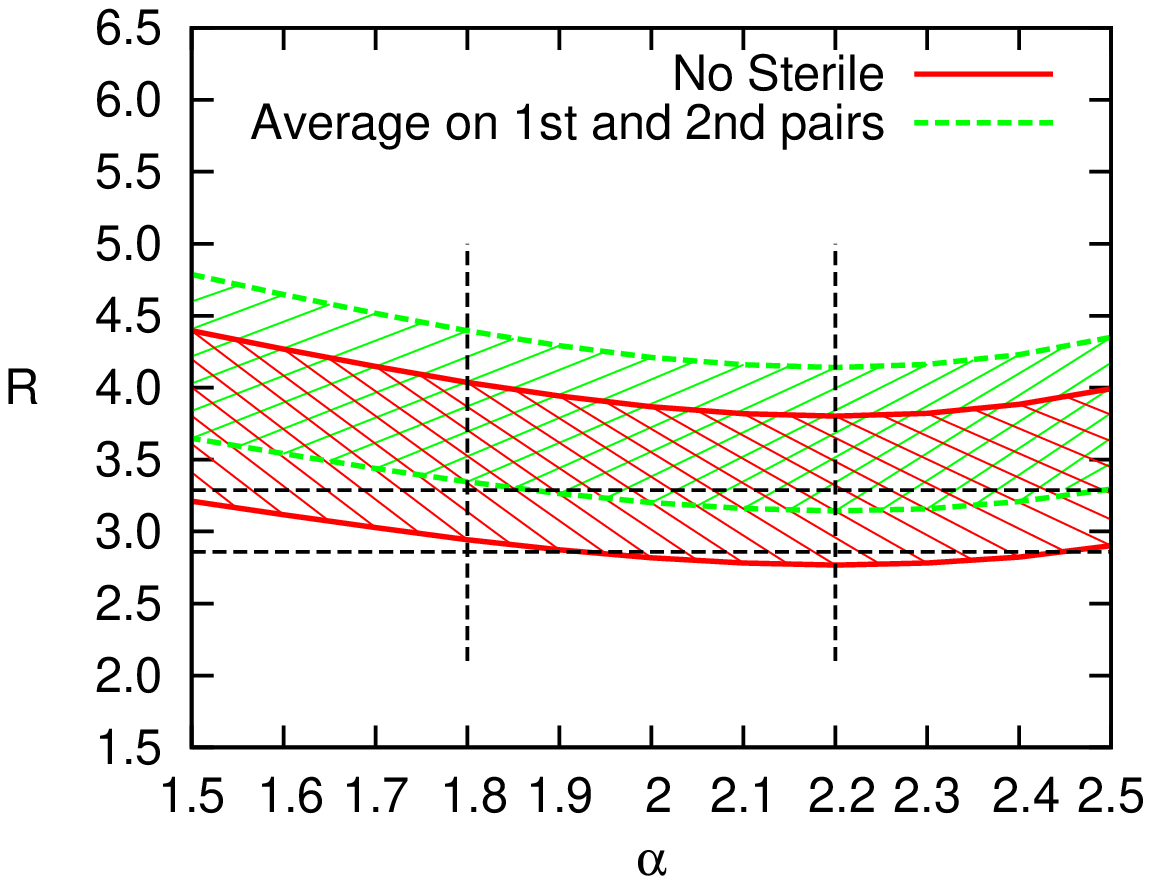}}
 \centerline{\hspace{0.45cm}(c)\hspace{8.45cm}(d)}
 \centerline{\vspace{-2.0cm}}
 \centerline{\includegraphics[bb=260 70 570
 500,keepaspectratio=true,clip=true,angle=-90,scale=0.6]{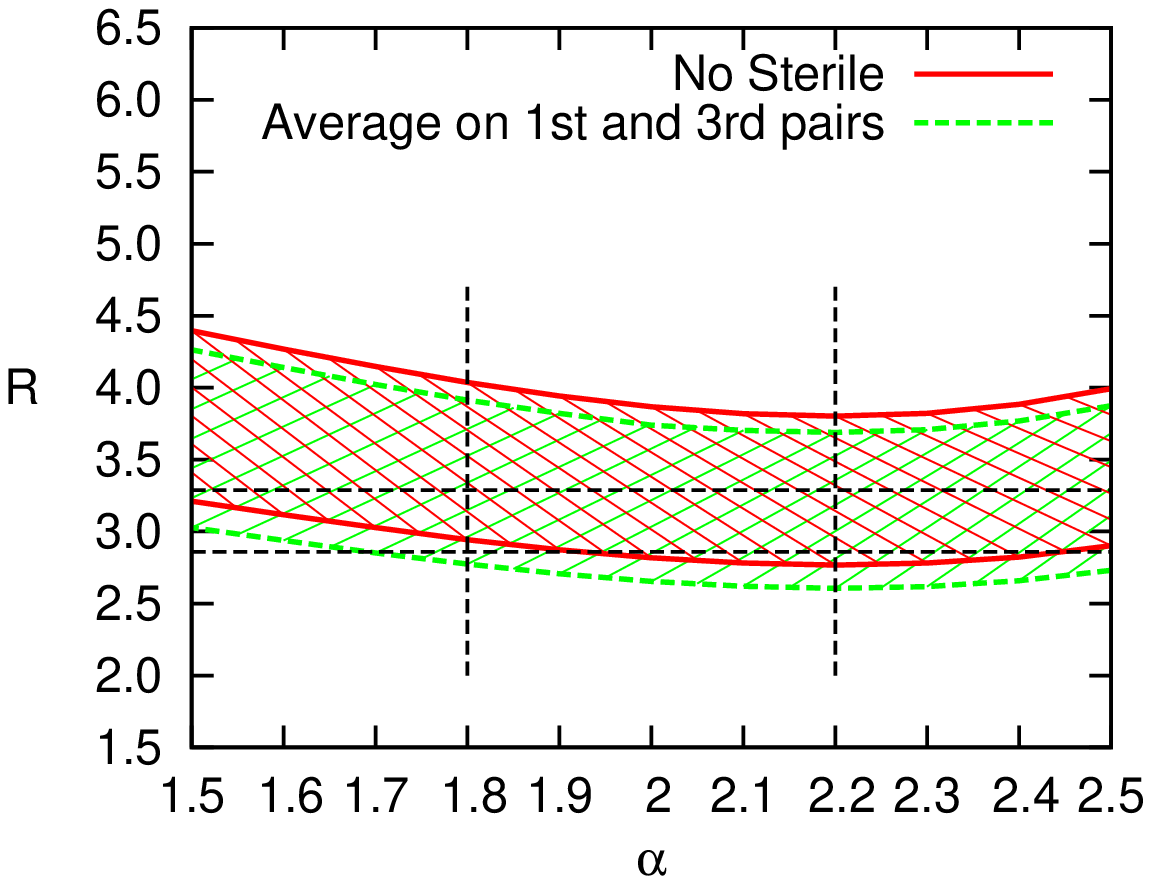}\includegraphics[bb=260 70 570
 450,keepaspectratio=true,clip=true,angle=-90,scale=0.6]{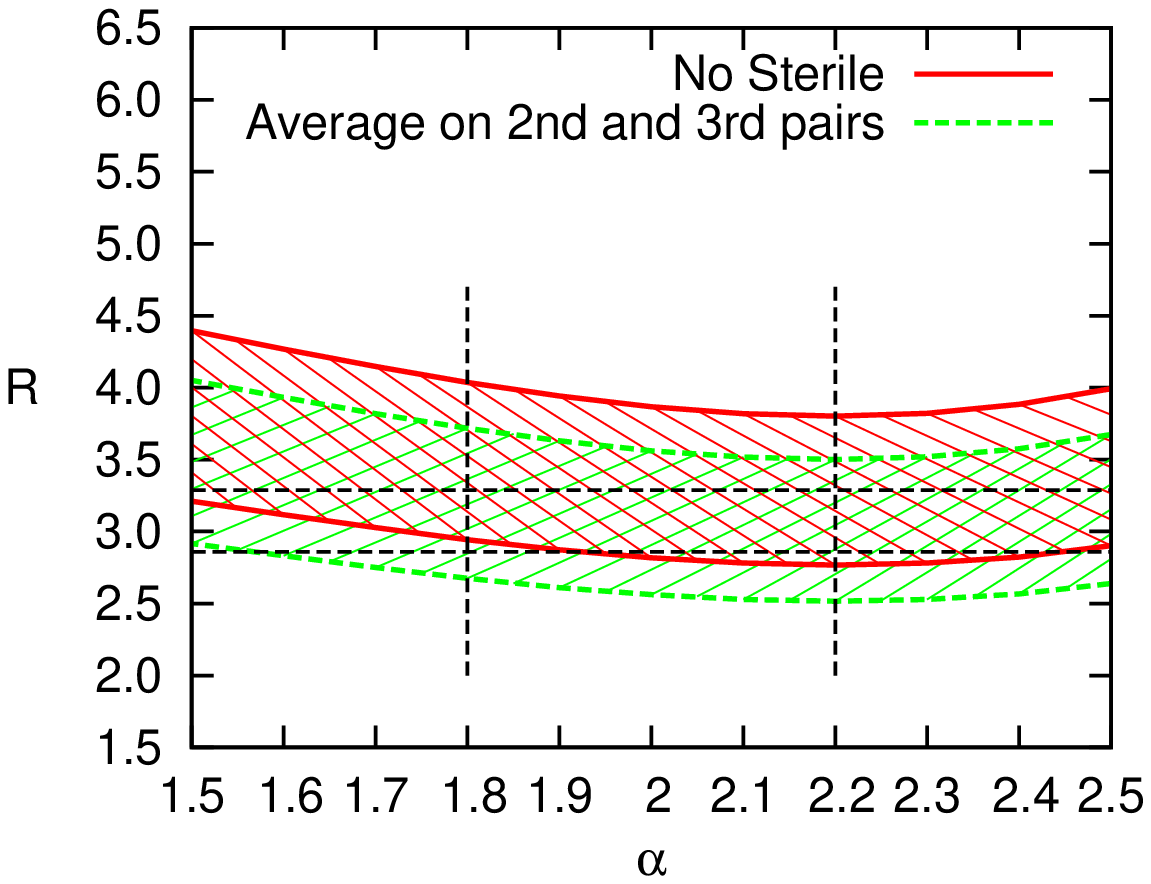}}
 \centerline{\hspace{0.45cm}(e)\hspace{8.45cm}(f)}
 \centerline{\vspace{-1.7cm}}
 \end{center}
 \caption{{\small The same as Fig.~(\ref{010}) except that the
 input parameters have been varied in the future uncertainty
 intervals of Table~\ref{para}. Particularly, we assumed 5~\%
 uncertainty interval for the 13-mixing angle with the central
 value near the present upper bound: $\sin^2\theta_{13}=0.03$.
 } }
  \label{010new}
\end{figure}

Figs.~(\ref{010})~and~(\ref{010new}) show the dependence of $R$ on
$\alpha$ for the stopped-muon source; where $\bar{R}_\mu=3.07$.
The input parameters have been varied in the current uncertainty
intervals in Fig.~(\ref{010}). Comparing Fig.~(\ref{010}) with
Fig.~(\ref{120}), it is obvious that the hatched areas are wider
for the stopped-muon source, which means that the recognition of
sterile neutrinos is harder for this kind of sources. In contrast
to the pion source, reducing the uncertainty of input parameters
to the future uncertainty intervals do not lead to a separation
between the hatched areas in Fig.~(\ref{010new}). However, some
cases can marginally be discriminated such as the case shown in
Fig.~(\ref{010new}-d).

\begin{figure}[t]
  \begin{center}
 \centerline{\includegraphics[bb=310 -100 570
 520,keepaspectratio=true,clip=true,angle=270,scale=0.68]{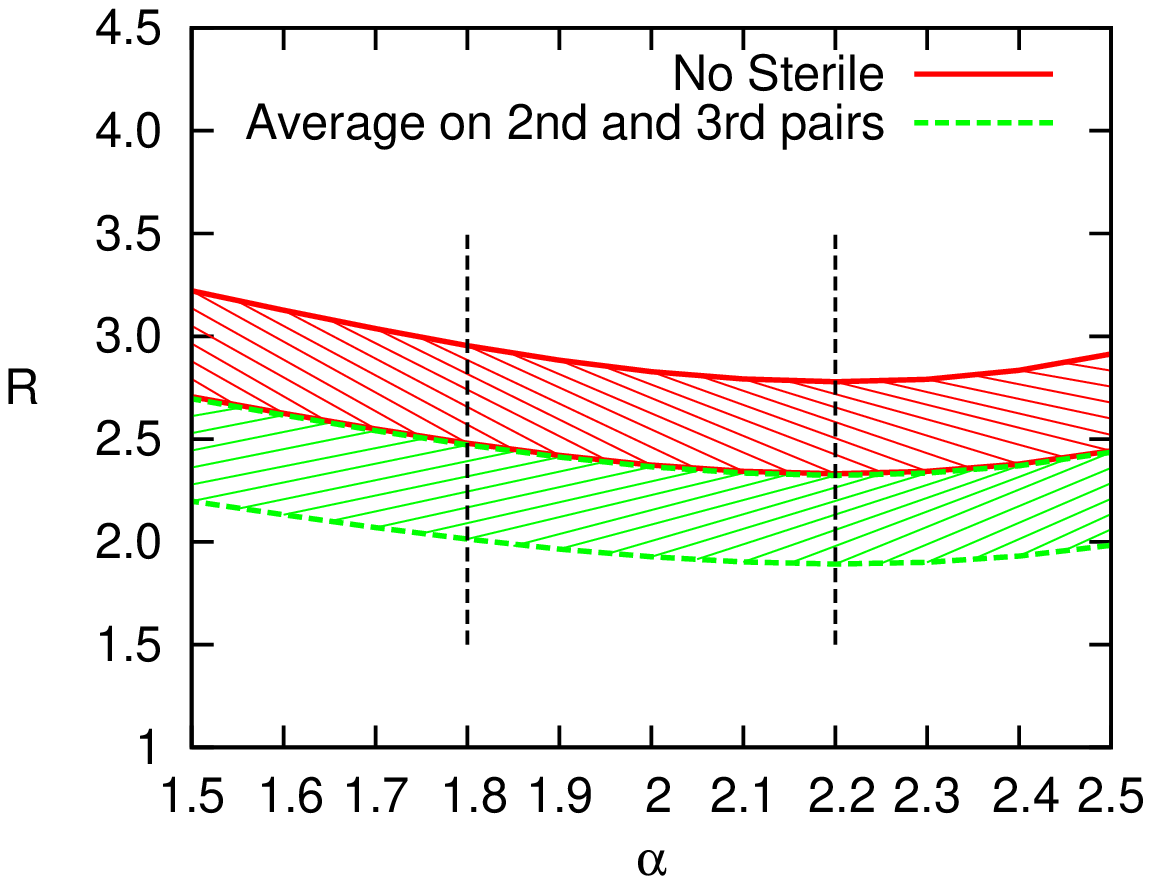}
 \includegraphics[bb=0 -20 290
 470,clip=true,angle=270,width=10cm]{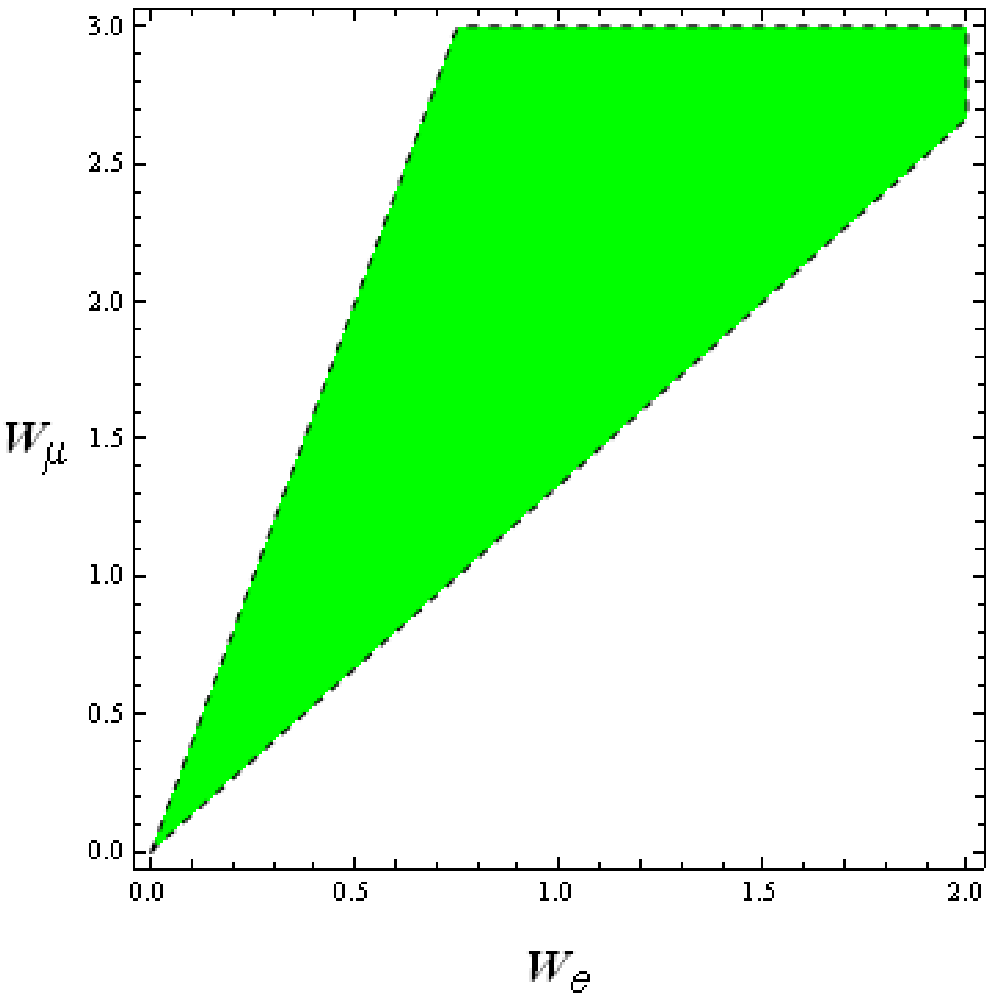}}
 \centerline{\hspace{2.5cm}(a)\hspace{9.45cm}(b)}
 \centerline{\vspace{-2.0cm}}
 \end{center}
 \caption{{\small The hatched areas in (a) show the dependence of
 $R$ on $\alpha$ for the case of no sterile neutrinos (red) and
 the case with
 average on the 2nd and 3rd pairs of almost degenerate neutrinos
 (green). The initial flavor ratio $w_e:w_\mu:0$ has been varied
 in the region depicted in (b). In drawing this figure we have assumed
 the best-fit values for the mixing parameters and varied $\lambda_e\in(0.8,1)$.
 } }
  \label{robustness}
\end{figure}

In drawing Figs.~(\ref{120},\ref{120new}) and
Figs.~(\ref{010},\ref{010new}) we have assumed initial flavor
ratios $1:1.85:0$ and $0:1:0$, respectively. However, it should be
noticed that the initial flavor ratio can deviates from these
values due to the interplay of different mechanisms in neutrino
production or effects that have not been considered in the
calculation of these values. For example, in the pion source, a
{\em part} of the produced muons in the decays of $\pi^\pm$ (not
all of them) can lose their energy before decay such that the
initial flavor ratio of neutrinos takes a value between the two
extreme cases of pion and stopped-muon sources. The question that
arises here is that to what extent the results of this section
(for example the separation between hatched areas in
Fig.~(\ref{120new}-f)) are robust against the deviations of the
initial flavor ratio $w_e:w_\mu:w_\tau$ from the assumed values
\footnote{Conversely, by assuming standard propagation of
neutrinos between the source and detector, measurements of flavor
ratios at Earth can be used to set bounds on the initial flavor
ratios at the source \cite{Esmaili,Choubey:2009jq}.}. To answer
this question we consider particularly the case of
Fig.~(\ref{120new}-f) where the hatched areas are completely
separated. Also we use the following parametrization of the
initial flavor ratio \footnote{a slightly different
parametrization $w_e:w_\mu:w_\tau=1:n:0$ has been used in
\cite{Choubey:2009jq}.}
\begin{equation} \label{initial} w_e:w_\mu:w_\tau=n:1:0.
\end{equation} The $w_\tau=0$ in the above parametrization comes
from the fact that the number of prompt $\nu_\tau$ (from the
decays $D_s\to\tau\nu_\tau,\ldots$) is very small and can be
neglected. In Fig.~(\ref{robustness}-a) we have varied $n$ in
Eq.~(\ref{initial}) such that the separated regions in
Fig.~(\ref{120new}-f) begin to overlap. In drawing
Fig.~(\ref{robustness}-a) we have varied $n\in(0.25,0.75)$. The
initial flavor ratios $w_e:w_\mu:0$ corresponding to this interval
are shown in Fig.~(\ref{robustness}-b). As can be seen, for nearly
large deviations of the initial flavor ratio from $1:1.85:0$, the
hatched areas remain separated, which means that the lack of
knowledge about the exact value of the initial flavor ratio of
neutrinos at the source do not affect substantially the capability
of neutrino telescopes in probing pseudo-Dirac neutrino scenario.

\section{Conclusion\label{conclusion}}

The new generation of km$^3$ neutrino telescopes opens a new
window to study cosmos via the detection of high energy neutrinos
predicted to be emitted from the astrophysical objects. Because of
the extremely large distance of the sources ($\gtrsim$ Mpc) the
flavor oscillation is sensitive to the very tiny mass squared
differences $10^{-18}$~eV$^2$~$\lesssim\Delta
m^2\lesssim10^{-12}$~eV$^2$. The existence of sterile neutrinos
with masses almost degenerate with the active ones such that the
mass squared differences between the active and sterile neutrinos
lie in the above region is hypothesized in many models (the
so-called pseudo-Dirac scenario.) We have studied the effect of
these sterile neutrinos on the flux of cosmic neutrinos and
discussed the capability of IceCube to verify the existence of
them. In the analysis we have considered different cases
corresponding to the existence of sterile neutrinos degenerate in
mass with different $\nu_{iL}$ such that the oscillatory terms
driven by these mass squared differences can be averaged out.

The detection power of IceCube (and other proposed km$^3$ neutrino
telescopes), in the range of neutrino energies 100~GeV~$<E_\nu<
100$~TeV, is limited to distinguishing two types of events:
$\mu$-track and shower-like events. We have considered the ratio
of these events, $R$ in Eq.~(\ref{ratioR}), as the realistic
quantity that can be measured in the IceCube. We have studied the
possibility of using the measured value of $R$ as a discriminator
between the pseudo-Dirac scenario and the scenario with no sterile
neutrinos. We have considered various sources of uncertainties in
our analysis. One part of these uncertainties comes from the
imprecision of the neutrino oscillation experiments in the
determination of mixing parameters. For this part we considered
two sets of uncertainty intervals for the mixing parameters: i)
the current uncertainty intervals from the performed experiments;
and ii) the future uncertainty intervals that will be achieved in
the forthcoming experiments. Both of these sets have been depicted
in Table~\ref{para}. Among the mixing parameters, the measurable
quantity $R$ is most sensitive to the exact value of the mixing
angle $\theta_{23}$. For the uncertainty of this parameter we have
used the 6~\% precision on $\sin^2\theta_{23}$ \cite{:2008ee}
depicted in Table.~\ref{para}. The other part of uncertainties
comes from the not completely known mechanism of neutrino
production in the sources. Many models of neutrino production
predict a power-law spectrum for the cosmic neutrinos. However,
the value of the spectral index in the power-law spectrum depends
on the details of the neutrino production mechanism and can take
values in the interval $(1,3)$. Also, the ratio of the number of
electron anti-neutrinos to the number of electron neutrinos is not
known. We have taken into account all these uncertainties in our
analysis.

The analysis has been done for two different initial flavor
composition $w_e:w_\mu:w_\tau$ at the source: i) pion source
$1:1.85:0$; and ii) stopped-muon source $0:1:0$. It has been shown
that with a higher precision of the mixing angles achievable in
the forthcoming oscillation experiments, in certain cases it is
possible to demonstrate or rule out the existence of sterile
neutrinos hypothesized in the pseudo-Dirac scenario. In these
cases the regions corresponding to existence and absence of
sterile neutrinos are well-separated in the parameter space for
neutrinos coming from pion sources. For example, it is very
promising to probe the case of two sterile neutrinos with masses
almost degenerate with $\nu_{2L}$ and $\nu_{3L}$ (see
Fig.~(\ref{120new}-f)). For neutrinos coming from stopped-muon
sources, these regions mostly overlap such that their
discrimination cannot be done without ambiguity. Also, the
robustness of these results has been tested against the
uncertainties in the initial flavor ratio of neutrinos at the
source. It has been shown that for reasonably large variations of
the initial flavor ratio around the expected value for pion
source, $1:1.85:0$ (see Fig.~(\ref{robustness}) for
clarification), the regions corresponding to the existence and the
absence of sterile neutrinos remain separated.

\section*{Acknowledgement}

The author is grateful to Y.~Farzan for useful discussions and for
her valuable comments on the manuscript. Also, I would like to
thank H.~Firouzjahi for the careful reading of the manuscript and
valuable comments. I would like to thank ``Bonyad-e Melli-e
Nokhbegan'' for partial financial support.

\end{document}